\documentclass[aps,pra,reprint,showpacs,floatfix]{revtex4-1}

\usepackage[utf8]{inputenc}
\usepackage[T1]{fontenc}
\usepackage[english]{babel}

\usepackage{graphicx}
\usepackage{subfigure}

\usepackage{bm}

\usepackage{units}
\usepackage[squaren]{SIunits}

\usepackage[version=3]{mhchem}

\usepackage{hyperref}

\graphicspath{{./pics/}}

\newcommand{\ee}{\ensuremath{\mathrm{e}}} 
\newcommand{\ii}{\ensuremath{\mathrm{i}}} 

\newcommand{\abs}[1]{\ensuremath{\left\vert #1 \right\vert}} 
 
\newcommand{\eto}[1]{\ensuremath{\,{\ee}^{#1}}} 
\renewcommand{\vec}[1]{{\ensuremath{\bm{#1}}}} 
\newcommand{\dd}{\ensuremath{\mathrm{d}}} 
 
\newcommand{\firstderiv}[2]{\ensuremath{\frac{\mathrm{d}#1}{\mathrm{d}#2}}} 
\newcommand{\secondderiv}[2]{\ensuremath{\frac{\mathrm{d}^2#1}{\mathrm{d}{#2}^2}}}	
\newcommand{\firstpderiv}[2]{\ensuremath{\frac{\partial #1}{\partial #2}}} 
 
\newcommand{\qmprod}[2]{\ensuremath{\left\langle #1\middle\vert #2 \right\rangle}} 
 
\DeclareMathOperator{\Real}{Re} 
 
\begin{document}

\title{Relation between the eigenfrequencies of Bogoliubov excitations
  of Bose-Einstein condensates and the eigenvalues of the Jacobian in
  a time-dependent variational approach}
\date{\today}
\author{Manuel Kreibich}
\author{J\"org Main}
\author{G\"unter Wunner}
\affiliation{1. Institut f\"ur
  Theoretische Physik, Universit\"at Stuttgart, 70550 Stuttgart,
  Germany}

\begin{abstract}
  We study the relation between the eigenfrequencies of the Bogoliubov
  excitations of Bose-Einstein condensates, and the eigenvalues of the
  Jacobian stability matrix in a variational approach which maps the
  Gross-Pitaevskii equation to a system of equations of motion for the
  variational parameters. We do this for Bose-Einstein condensates
  with attractive contact interaction in an external trap, and for a
  simple model of a self-trapped Bose-Einstein condensate with
  attractive $1/r$ interaction. The stationary solutions of the
  Gross-Pitaevskii equation and Bogoliubov excitations are calculated
  using a finite-difference scheme. The Bogoliubov spectra of the
  ground and excited state of the self-trapped monopolar condensate
  exhibits a Rydberg-like structure, which can be explained by means
  of a quantum defect theory. On the variational side, we treat the
  problem using an ansatz of time-dependent coupled Gaussians combined
  with spherical harmonics. We first apply this ansatz to a condensate
  in an external trap without long-range interaction, and calculate
  the excitation spectrum with the help of the time-dependent
  variational principle. Comparing with the full-numerical results, we
  find a good agreement for the eigenfrequencies of the lowest
  excitation modes with arbitrary angular momenta. The variational
  method is then applied to calculate the excitations of the
  self-trapped monopolar condensates, and the eigenfrequencies of the
  excitation modes are compared.
\end{abstract}


\maketitle

\section{Introduction}
\label{sec:introduction}
In the quantum mechanical description of the ground states of
Bose-Einstein condensates in the framework of the Gross-Pitaevskii
equation, the frequencies of elementary excitations of the condensates
are obtained by solving the Bogoliubov-de Gennes equations.  In an
alternative description, a variational approach with coupled Gaussian
functions has recently been proposed by Rau et
al. \cite{Rau10a,Rau10b} which maps the Gross-Pitaevskii equation to a
dynamical system for the variational parameters that can be analyzed
using the familiar tools of classical nonlinear dynamics. Ground
states correspond to the fixed points of the dynamical system, and
their stability properties follow from the eigenvalues of the Jacobian
at the fixed points.  In this paper we shall investigate the question
whether or not there is a relation between the eigenvalues of the
Jacobian and the eigenfrequencies of the quantum mechanical Bogoliubov
excitations, and if so, to what extent the eigenvalues of the Jacobian
can reproduce the frequencies of these excitations.

The realization of a Bose-Einstein condensate (BEC) with \ce{^{52}Cr}
atoms \cite{Griesmaier05a} marked the beginning of experimental
investigations of BECs with long-range interactions. The anisotropic
dipole-dipole interaction caused by the large magnetic moment of the
\ce{^{52}Cr} atoms influences the properties of the quantum gas
\cite{Stuhler05a}, and is responsible for new phenomena, such as a
roton-maxon spectrum \cite{Santos03a}, structured ground states
\cite{Ronen07a,Dutta07a}, and angular collapse \cite{Wilson09a}.
Recently a condensate of \ce{^{164}Dy} atoms with an even larger
magnetic moment was created \cite{Lu10a,Lu11a}, and BECs of other
lanthanides with a strong dipole-dipole interaction should be possible
\cite{McLeland06a}.

A model of a BEC with a different long-range interaction was proposed
by O'Dell et al. \cite{ODell00a}. In contrast to the dipolar
interaction, this interaction is monopolar, i.e., ``gravity-like''
with an attractive $1/r$ potential. Although it will be difficult to
realize this model experimentally, BECs with monopolar long-range
interaction are worth investigating in their own right, since they
exhibit the phenomenon of self-trapping \cite{ODell00a}, i.e., the
existence of a stable condensate without an additional external trap.
Furthermore, the isotropic character of the interaction renders
numerical investigations easier than in the anisotropic case, and
therefore BECs with monopolar interaction can serve as model systems
for the treatment of condensates with long-range interactions to test
new approaches and techniques.

The stationary states of self-trapped monopolar condensates have been
calculated in the Thomas-Fermi regime and with the variational ansatz
of a single Gaussian \cite{ODell00a}, full-numerically
\cite{Papadopoulos07a}, and with an ansatz of coupled Gaussians
\cite{Rau10a,Rau10b}. Several aspects of the excitation spectrum have
also been investigated \cite{Giovanazzi01a,Cartarius08b,Rau10b}, but a
comprehensive study is still lacking. In this paper we will solve the
Bogoliubov-de Gennes equations and reveal a Rydberg-like structure in
the numerically exact Bogoliubov spectra, similar to the spectra of
alkali metals.

The full-numerical calculations are very accurate, if -- depending on
the method -- grid size, number of basis functions, etc., are chosen
carefully, but may become computationally very expensive.  As an
alternative we pursue a variational ansatz with coupled Gaussian
functions. Single Gaussians have been used in the literature to obtain
qualitative results for BECs (e.g.\ in
\cite{ODell00a,Perez-Garcia96a}). The ansatz can be extended to
time-dependent coupled Gaussians \cite{Heller76a,Heller81a}, and it
was demonstrated \cite{Rau10a,Rau10b} that the method can
quantitatively reproduce the properties of the stationary solutions of
the Gross-Pitaevskii equation with both monopolar and dipolar
long-range interaction.  However, as we discuss below, the ansatz with
coupled Gaussians can only describe excitations with a maximum angular
momentum of $l=2$. Several extensions of a Gaussian ansatz have been
considered in the literature, e.g., Gaussians with Hermite or Laguerre
polynomials \cite{Ronen07a,Buccoliero07a,Buccoliero09a}, or sines and
cosines \cite{Maucher10a}. But these methods allow for no systematic
improvement of the ansatz, which is the case for the variational
method we present in this paper.

Our variational ansatz is based on a combination of coupled Gaussians
with spherical harmonics, and can describe excitations with arbitrary
angular momenta in radially symmetric systems. The power of the method
will be demonstrated by applying it to BECs without and with monopolar
long-range interaction.

The paper is organized as follows. In Sec.~\ref{sec:monopolar} we give
the basic equations, and describe our numerical method for calculating
the stationary states and excitations of self-trapped monopolar
condensates. We show that the Bogoliubov spectra can be nicely
analyzed in terms of quantum defect theory.  Our variational ansatz
with time-dependent coupled Gaussians combined with spherical
harmonics is described in Sec.~\ref{sec:vari-appr-with}, and the
equations of motion for the Gaussian parameters are derived.  The
method is applied to BECs without and with the monopolar long-range
interaction. In Sec.~\ref{sec:conclusion-outlook} we draw conclusions
and give an outlook on future work.

\section{Full-numerical treatment of the self-trapped monopolar
  condensate}
\label{sec:monopolar}

The time-dependent Gross-Pitaevskii equation (GPE) for the
self-trapped condensate with short-range contact interaction and
long-range monopolar interaction reads
\begin{align}
  \label{eq:gpemonotime}
  \ii \firstpderiv{\psi}{t} (\vec{r},t) = 
  \Biggl[
    &- \Delta
    + 8 \pi a \abs{\psi(\vec{r},t)}^2 \nonumber \\
    &- 2 \int \dd^3 r^\prime
    \frac{\abs{\psi(\vec{r}^\prime,t)}^2}{\abs{\vec{r}-\vec{r}^\prime}}
  \Biggr]
  \psi(\vec{r},t),
\end{align}
where $a$ denotes the s-wave scattering length. Since we will
concentrate on the case of self-trapping, the external potential has
been omitted. All variables in Eq.~(\ref{eq:gpemonotime}) are given in
the natural units introduced in \cite{Papadopoulos07a}: Lengths are
measured in units of the ``Bohr radius'' $a_\text{u} = \hbar^2/mu$,
energies in units of the ``Rydberg energy'' $E_\text{u} = u / 2
a_\text{u}$, and time in units of $t_\text{u} = \hbar/E_\text{u}$. The
quantity $u$ is the coupling constant of the monopolar interaction
defined in \cite{ODell00a} and depends on the intensity and wave
number of the laser, and the polarizability of the atoms.

Eq.~(\ref{eq:gpemonotime}) represents the GPE for the fictitious
one-boson problem. One can make use of the scaling property of
\cite{Papadopoulos07a} to scale all quantities to an $N$-boson system:
\begin{align}
  (\vec{r}, a, t, \psi) \to
  (N \vec{r}, N^2 a, N^2 t, N^{-3/2} \psi).
\end{align}
The scaled dimensionless units are used throughout this work and in
all figures whenever considering monopolar condensates. In these
units, the only remaining parameter is the scattering length $a$
\cite{Papadopoulos07a}. The stationary GPE can be obtained by
substituting $\psi(\vec{r},t) = \psi (\vec{r}) \exp(-\ii \mu t)$, with
the chemical potential $\mu$, in the time-dependent GPE
(\ref{eq:gpemonotime}), which leads to
\begin{align}
  \label{eq:gpemonostat}
  \mu \psi (\vec{r}) = 
  \left[
    - \Delta
    + 8 \pi a \abs{\psi(\vec{r})}^2 
    - 2 \int \dd^3 r^\prime
    \frac{\abs{\psi(\vec{r}^\prime)}^2}{\abs{\vec{r}-\vec{r}^\prime}}
  \right]
  \psi(\vec{r}).
\end{align}

\subsection{Calculation of stationary solutions}
\label{sec:stationary-solutions}

For a numerical treatment of the stationary GPE (\ref{eq:gpemonostat})
it is convenient to convert the integro-differential equation into two
coupled differential equations. This can be achieved by defining the
mean-field potential
\begin{align}
  \label{eq:monomeanfield}
  \phi(\vec{r}) = - 2 \int \dd^3 r^\prime
  \frac{\abs{\psi(\vec{r}^\prime)}^2}{\abs{\vec{r}-\vec{r}^\prime}}.
\end{align}
Since we search for radially symmetric stationary solutions we assume
the wave function and the mean-field potential to depend only on the
radial coordinate: $\psi(\vec{r}) = \psi(r)$ and $\phi(\vec{r}) =
\phi(r)$. Letting the Laplacian in spherical coordinates act on
Eq.~(\ref{eq:monomeanfield}) one obtains the two one-dimensional,
nonlinear coupled differential equations
\begin{subequations}
  \label{eq:gpemonosys}
  \begin{align}
    \left(
      - \secondderiv{}{r}
      -\frac{2}{r} \firstderiv{}{r}
      + 8 \pi a \abs{\psi(r)}^2 
      + \phi(r)
    \right)
    \psi(r) &= \mu \psi(r), \\
    \left(
      \secondderiv{}{r}
      +\frac{2}{r} \firstderiv{}{r}
    \right)
    \phi(r)
    - 8 \pi \abs{\psi(r)}^2 &= 0.
  \end{align}
\end{subequations}

The system of Eqs.~(\ref{eq:gpemonosys}) can be solved numerically 
in different ways.
Since it is a one-dimensional problem,
one can integrate the equations using a Runge-Kutta algorithm from
$r=0$ to a sufficiently large value $r_\text{max}$ with appropriately
chosen initial conditions for $\psi(0)$, $\psi^\prime(0)$, $\phi(0)$
and $\phi^\prime(0)$ \cite{Papadopoulos07a, Cartarius08a,
  Cartarius08b}. Their values must be varied until the wave function
converges towards zero at $r = r_\text{max}$. With this method the
ground and excited state can be calculated efficiently. However, to
obtain a normalized solution $\psi(r)$ the
wave function, scattering length, and mean field energy must be
rescaled. Thus, it is difficult to obtain a solution for a  given 
fixed value of the scattering length. Additionally, it is not easy to calculate
the Bogoliubov spectrum of the system with this method, since the
solutions of the Bogoliubov-de Gennes (BDG) equations have large
extensions, and a very big value of $r_\text{max}$ has to be
chosen. For example, to calculate 20 eigenvalues for an angular
momentum of $l=6$, $r_\text{max}$ needs to be larger than $1000$. In
this case, machine precision in the Runge-Kutta method is not
sufficient to obtain converged solutions, leaving this method useless
for higher modes. In \cite{Cartarius08b, Rau10b}, only the three
lowest $l=0$ modes could be calculated.

Another method is the imaginary time evolution (replacement $t \to t =
\ii \tau$ in Eq.~(\ref{eq:gpemonotime})) of an initial wave function
on a grid. As time evolves the wave function converges to the ground
state. This method is useful to find the ground state or a metastable
state of a system. However, a collectively excited state, as we
consider below, cannot be obtained by imaginary time evolution.

To avoid these disadvantages, we use the finite-difference method to
solve the stationary GPE (\ref{eq:gpemonosys}): Wave functions and the
mean-field potential are discretized on a grid and all derivatives are
replaced by their finite-difference approximation. To arrive at a
closed system of algebraic equations which can be solved by a
nonlinear root search one needs appropriate boundary conditions:
$\psi^\prime(0) = 0$ and $\phi^\prime(0) = 0$, to ensure that the
functions are differentiable at the origin, and $\psi(r_\text{max}) =
0$ to obtain a normalizable wave function. The fourth boundary
condition can be obtained by looking at the asymptotic behavior of the
mean-field potential (\ref{eq:monomeanfield}). Approximating
$1/\abs{\vec{r}-\vec{r}^\prime} \approx 1/r$ for $r \gg r^\prime$ and
assuming a normalized wave function $\psi$, one obtains from
Eq.~(\ref{eq:monomeanfield}) $\phi(r) \approx -2/r$. The fourth
boundary condition is therefore set to be $\phi(r_\text{max}) =
-2/r_\text{max}$.

We perform the nonlinear root search using the Powell hybrid method
\cite{Powell70a}. In addition to the equations originating from the
finite-difference scheme, the normalization condition has to be
included, as well as the chemical potential as a parameter to be
determined by the root search.

\subsection{Bogoliubov-de Gennes equations}
\label{sec:bogoliubov-de-gennes}

The stability and elementary excitations of a self-trapped monopolar
condensate have already been analyzed in the literature: the lowest
monopole and quadrupole oscillation analytically and numerically
\cite{Giovanazzi01a}, the first monopole modes \cite{Cartarius08a},
and the lowest monopole and quadrupole modes by means of a variational
ansatz with coupled Gaussians \cite{Rau10b}. However, to the best of
our knowledge, a calculation of the Bogoliubov spectrum for arbitrary
angular momenta and higher excitations does not yet exist.

To derive the BDG equations, one starts from the usual ansatz for a
perturbation of a stationary state
\begin{align}
  \label{eq:bdgansatz}
  \psi(\vec{r},t) =
  \left[
    \psi_0(\vec{r})
    + \lambda
    \left(
      u(\vec{r}) \eto{-\ii \omega t}
      + v^*(\vec{r}) \eto{\ii \omega t}
    \right)
  \right]
  \eto{-\ii \mu t},
\end{align}
where $\omega$ is the frequency and $\lambda$ the amplitude of the
perturbation ($\abs{\lambda} \ll 1$), and $\mu$ is the chemical
potential of the stationary solution $\psi_0$ with corresponding
mean-field potential $\phi_0$. Eq.~(\ref{eq:bdgansatz}) is inserted
into the time-dependent GPE (\ref{eq:gpemonotime}), terms of second
order in $\lambda$ are neglected, and collecting terms evolving in
time with $\exp(-\ii \omega t)$ and $\exp(\ii \omega t)$ yields the
BDG equations
\begin{widetext}
  \begin{subequations}
    \label{eq:bdgmono}
    \begin{align}
      \omega u(\vec{r}) &=
      \left[
        - \Delta
        - \mu
        + 16 \pi a \abs{\psi_0(\vec{r})}^2
        + \phi_0(\vec{r})
      \right]
      u(\vec{r})
      + 8 \pi a (\psi_0(\vec{r}))^2 v(\vec{r})
      + \psi_0(\vec{r}) f(\vec{r}), \\
      -\omega v(\vec{r}) &=
      \left[
        - \Delta
        - \mu
        + 16 \pi a \abs{\psi_0(\vec{r})}^2
        + \phi_0(\vec{r})
      \right]
      v(\vec{r})
      + 8 \pi a (\psi_0^*(\vec{r}))^2 u(\vec{r})
      + \psi_0^*(\vec{r}) f(\vec{r}),
    \end{align}
  \end{subequations}
\end{widetext}
with the auxiliary field (similar to the mean-field potential)
\begin{align}
  \label{eq:bdgmonoaux}
  f(\vec{r}) =
  - 2 \int \dd^3 r^\prime
      \frac{\psi_0^*(\vec{r}^\prime) u(\vec{r}^\prime)
        + \psi_0(\vec{r}^\prime) v(\vec{r}^\prime)}
      {\abs{\vec{r}-\vec{r}^\prime}}.
\end{align}

The ansatz of Eq.~(\ref{eq:bdgansatz}) possesses a symmetry: the
exchange of $u(\vec{r}) \leftrightarrow v^*(\vec{r})$ and $\omega
\leftrightarrow -\omega$ leaves the ansatz invariant. Thus for each
solution $(u,v)$ and $\omega$ of Eqs.~(\ref{eq:bdgmono}), $(v^*,u^*)$
with $-\omega$ is another solution and both solutions represent the
same physical motion. For that reason, only solutions with $\Real
\omega \geq 0$ need to be considered. There are two solutions of
Eqs.~(\ref{eq:bdgmono}) which deserve special attention. If $\psi_0$
is assumed to be real, then $u(\vec{r}) = -v(\vec{r}) =
\psi_0(\vec{r})$ is a solution of Eqs.~(\ref{eq:bdgmono}) with the
frequency $\omega = 0$. This represents the well-known 
gauge transformation of the
condensate wave function $\psi(\vec{r}) \to \psi(\vec{r}) \exp(\ii
\phi)$ with a real phase $\phi$. This gauge mode does not describe a
physical motion of the condensate, and since it is always part of the
Bogoliubov spectrum, we will not discuss it when presenting the
results.

Furthermore, there always exist solutions of the BDG equations with
frequencies identical to the trapping frequencies
\cite{Pitaevskii03a}. These modes represent the center-of-mass
oscillations of the condensate along the three space directions with
angular momentum $l=1$. In the case of the self-trapped monopolar
condensate, there are no external traps and therefore the
frequencies are $\omega = 0$, which corresponds to a constant displacement
of the condensate.

Since the wave function $\psi_0$ and the mean-field potential $\phi_0$
are radially symmetric, we can separate the solutions $u$ and $v$ by
means of spherical harmonics
\begin{subequations}
  \label{eq:bdgmonosep}
  \begin{align}
    u_{nlm}(\vec{r}) &= Y_{lm}(\theta,\phi) u_{nl}(r), \\
    v_{nlm}(\vec{r}) &= Y_{lm}(\theta,\phi) v_{nl}(r),
  \end{align}
\end{subequations}
with the radial (excitation) quantum number $n$ and the usual angular momentum
quantum numbers $l,m$. Using the multipole expansion of the
integration kernel $1/\abs{\vec{r}-\vec{r}^\prime}$ (see, e.g.,
\cite{Arfken05a} and Eq.~(\ref{eq:multipole})), we can also express
the auxiliary field (\ref{eq:bdgmonoaux}) in the form
$f_{nlm}(\vec{r}) = Y_{lm}(\theta,\phi) f_{nl}(r)$ with ($\psi_0$ and
$\phi_0$ are assumed to be real from now on)
\begin{align}
  f_{nl}(r) =
  \frac{-8\pi}{2l+1}
  \int\limits_0^\infty \dd r^\prime \, (r^\prime)^2
  \frac{r_<^l}{r_>^{l+1}} \psi_0(r^\prime)
  [u_{nl}(r^\prime) + v_{nl}(r^\prime)],
\end{align}
where $r_< = \min(r,r^\prime)$ and $r_> = \max(r,r^\prime)$,
respectively. Inserting the Laplacian in spherical coordinates and
using the separation (\ref{eq:bdgmonosep}), we finally obtain from
Eqs.~(\ref{eq:bdgmono})
\begin{widetext}
  \begin{subequations}
    \label{eq:bdgmonosph}
    \begin{align}
      \omega_{nl} u_{nl}(r) &=
      \left[
        - \secondderiv{}{r}
        - \frac{2}{r} \firstderiv{}{r}
        + \frac{l(l+1)}{r^2}
        - \mu
        + 16 \pi a \psi_0^2(r)
        + \phi_0(r)
      \right]
      u_{nl}(r)
      + 8 \pi a \psi_0^2(r) v_{nl}(r)
      + \psi_0(r) f_{nl}(r), \\
      -\omega_{nl} v_{nl}(r) &=
      \left[
        - \secondderiv{}{r}
        - \frac{2}{r} \firstderiv{}{r}
        + \frac{l(l+1)}{r^2}
        - \mu
        + 16 \pi a \psi_0^2(r)
        + \phi_0(r)
      \right]
      v_{nl}(r)
      + 8 \pi a \psi_0^2(r) u_{nl}(r)
      + \psi_0(r) f_{nl}(r).
    \end{align}
  \end{subequations}
\end{widetext}

We solve Eqs.~(\ref{eq:bdgmonosph}) using the finite-difference
method. After choosing a grid, approximating the derivatives by finite
differences and replacing the integral in the auxiliary field $f$ by
an appropriate integration rule (we use the trapezoidal rule),
Eqs.~(\ref{eq:bdgmonosph}) turn into a matrix eigenvalue problem
\begin{align}
  \mathbf{M}
  \begin{pmatrix}
    u \\
    v
  \end{pmatrix}
  = \omega
  \begin{pmatrix}
    u \\
    v
  \end{pmatrix}.
\end{align}
The eigenvalues of the matrix $\mathbf{M}$ can then be found by
numerical diagonalization.

In actual calculations we found it advantageous to choose a
non-equidistant grid, since the solutions $u$ and $v$ can be highly
oscillatory near the origin, and at the same time extend to large
values of $r$. We use partially equidistant grids, i.e., an
equidistant grid with step size $\Delta r_1$ in the interval
$[0,r_1]$, another equidistant grid with a different $\Delta r_2$ in
the interval $[r_1,r_2]$, etc.

\subsection{Results}
\label{sec:results}

\begin{figure}
  \includegraphics[width=\columnwidth]{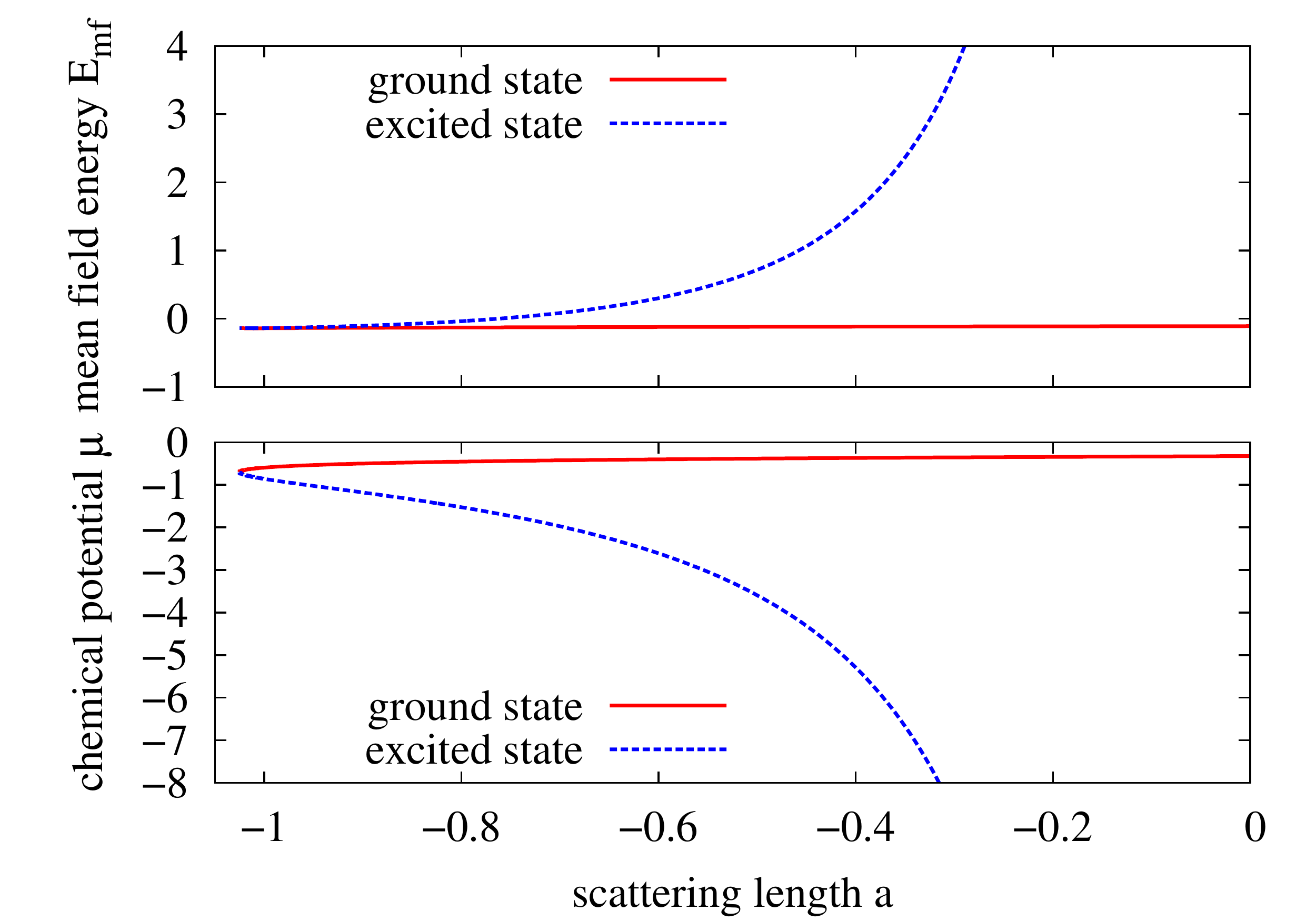}
  \caption{(Color online) Mean-field energy $E_\text{mf}$ and chemical
    potential $\mu$ of the ground and excited state of a self-trapped
    monopolar condensate as functions of the scattering length
    $a$. For a scattering length lower than the critical value of
    $a_\text{crit} \approx -1.025$ no stationary solution exists. At
    $a = a_\text{crit}$ the two solutions 
    emerge in a tangent bifurcation. For the ground state, both
    $E_\text{mf}$ and $\mu$ stay negative in the range of the
    scattering length considered. These
    quantities diverge for the excited state in the limit $a \to 0$.}
  \label{fig:mono_energies_num}
\end{figure}

Since the properties of the stationary solution have been discussed in
detail in the literature \cite{Papadopoulos07a, Cartarius08a,
  Cartarius08b}, we only give a brief review. Our results coincide
with those obtained using the outward integration method, and thus for
the stationary states both methods can be considered equally
applicable.  In Fig.~\ref{fig:mono_energies_num} we plot the
mean-field energy $E_\text{mf}$ and the chemical potential $\mu$ of
the ground and excited state as a function of the scattering length
$a$. Two solutions are born in a tangent bifurcation at the critical
scattering length $a = a_\text{crit} \approx -1.025$. At this point,
the mean-field energy, chemical potential and wave functions of the
ground and excited state merge. For $a \to 0$, the mean-field energy
and chemical potential of the excited state diverge, implying that
this state does not exist for $a \geq 0$.

\begin{figure}
  \includegraphics[width=\columnwidth]{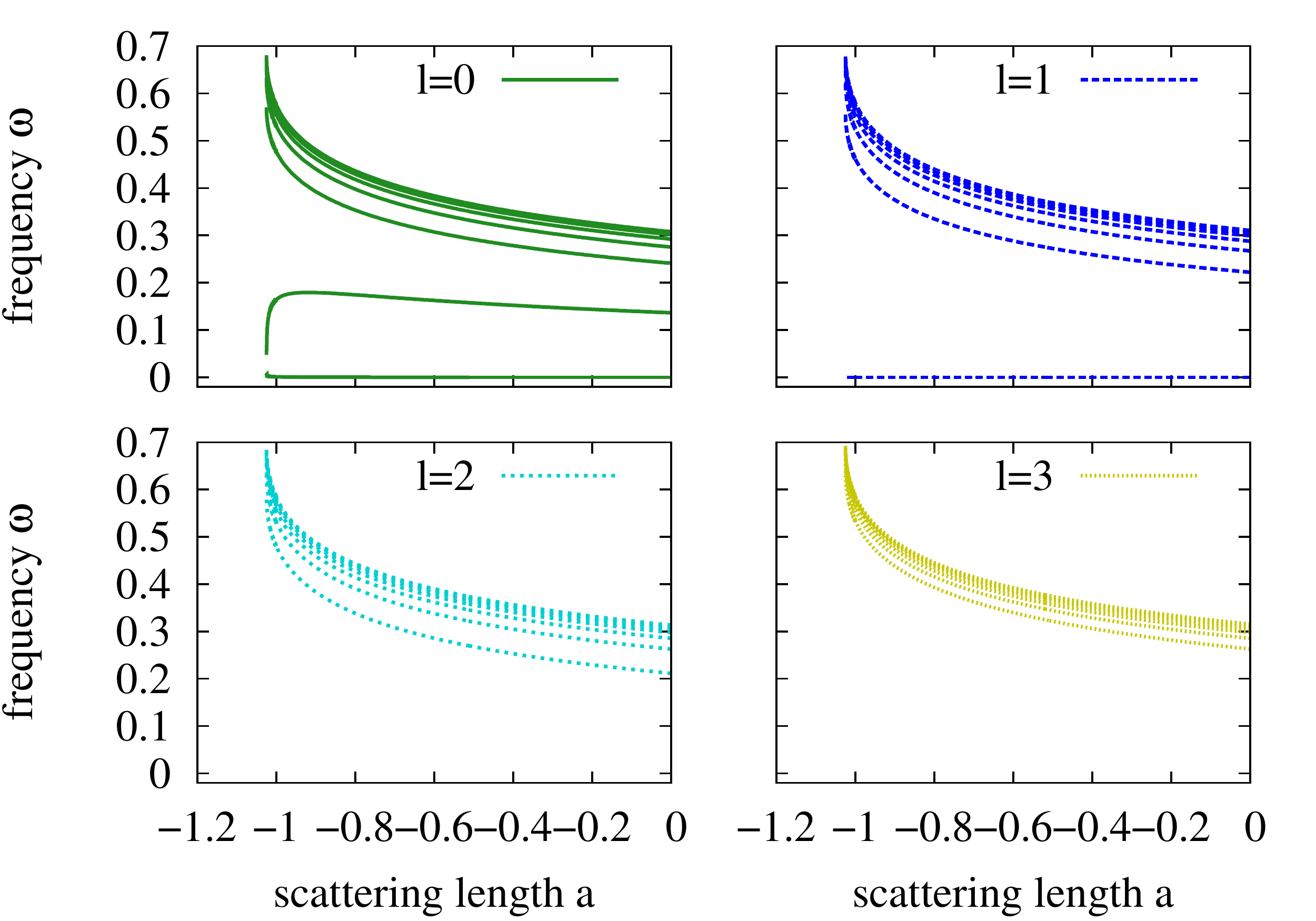}
  \caption{(Color online) Frequencies of Bogoliubov excitations of 
    the ground state
    in Fig.~\ref{fig:mono_energies_num} for the angular momenta from
    $l=0$ to $3$ as functions of the scattering length $a$.  The
    seven lowest eigenvalues are shown for each angular momentum. The
    spectrum only contains real frequencies, i.e., the ground state is
    stable. The lowest mode for $l=0$ tends to zero as $a \to
    a_\text{crit}$ which leads to the collapse of the condensate. The
    lowest $l=1$ mode corresponds to a displacement of the center-of-mass 
    of the condensate, while its shape remains unaffected. The
    frequency of this mode is exactly the trapping frequency
    \cite{Pitaevskii03a}, in this case $\omega = 0$. The frequencies
    of the other modes increase as the scattering length is decreased,
    finally merging with the modes of the excited state for $a \to
    a_\text{crit}$ (see Fig.~\ref{fig:mono_bogo_max}). Note that for fixed 
    scattering length the distance between
    adjacent frequencies diminishes with growing radial quantum
    number, indicating the convergence of the frequencies to a
    (scattering length dependent) limit frequency.}
  \label{fig:mono_bogo_min}
\end{figure}

Using the method described in Sec.~\ref{sec:bogoliubov-de-gennes} we
have calculated the Bogoliubov spectrum of the ground state.  For the
angular momenta from $l=0$ to $3$, Fig.~\ref{fig:mono_bogo_min} shows
the frequencies of the Bogoliubov excitations as a function of the
scattering length $a$. The ground state is stable, since its spectrum
contains only real frequencies. It can be seen that as the scattering
length is decreased towards its critical value the frequency of the
lowest mode with $l=0$ at first slightly increases but then goes to
zero at $a \to a_\text{crit}$, where the state vanishes. This mode is
responsible for the collapse of the condensate. The lowest $l=1$ mode
has the frequency $\omega = 0$ and corresponds to the displacement of
the center-of-mass of the condensate. This frequency remains
constantly $\omega =0$ as the scattering length is varied, and, as
already mentioned, corresponds to the (vanishing) trapping frequency.

\begin{figure}
  \includegraphics[width=\columnwidth]{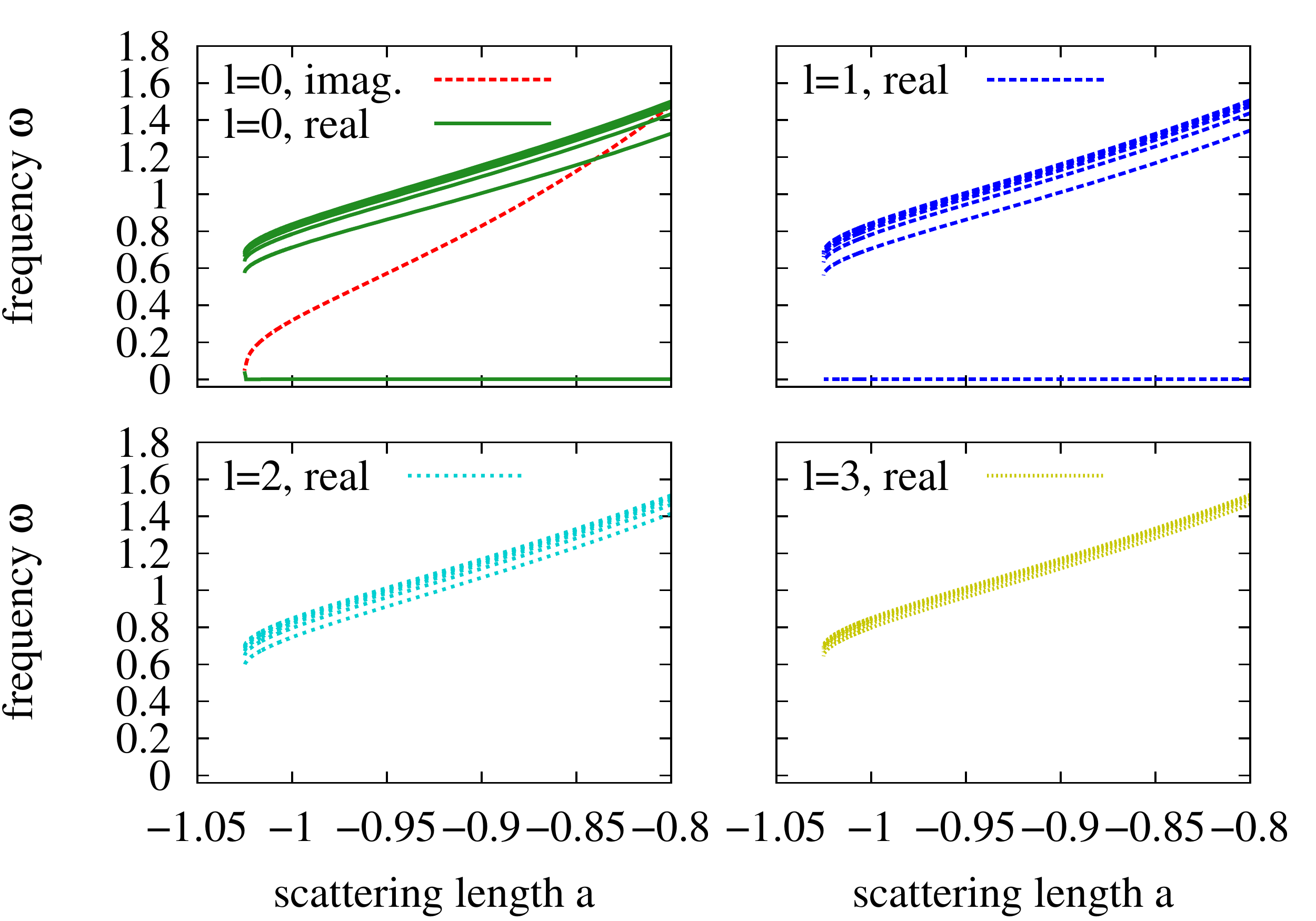}
  \caption{(Color online) Same as Fig.~\ref{fig:mono_bogo_min}, but
    for the excited state. There is one imaginary frequency for $l=0$:
    the excited state is unstable with respect to small
    perturbations. As for the ground state, the lowest $l=1$ mode is
    $\omega = 0$ and corresponds to a displacement of the
    center-of-mass of the condensate. Again, the frequencies of the
    stable modes apparently converge to a limit for fixed scattering
    length.}
  \label{fig:mono_bogo_max}
\end{figure}

The results for the excited state are presented in
Fig.~\ref{fig:mono_bogo_max}. All frequencies merge with those of the
ground state modes at the critical scattering length. There exists one
imaginary frequency for the angular momentum $l=0$. Therefore the
excited state is unstable with respect to this excitation, which leads
to a collapse with $l=0$ symmetry. As for the ground state the lowest
mode with $l=1$ represents the displacement of the condensate and is
constantly $\omega=0$.

\begin{figure}
  \includegraphics[width=\columnwidth]{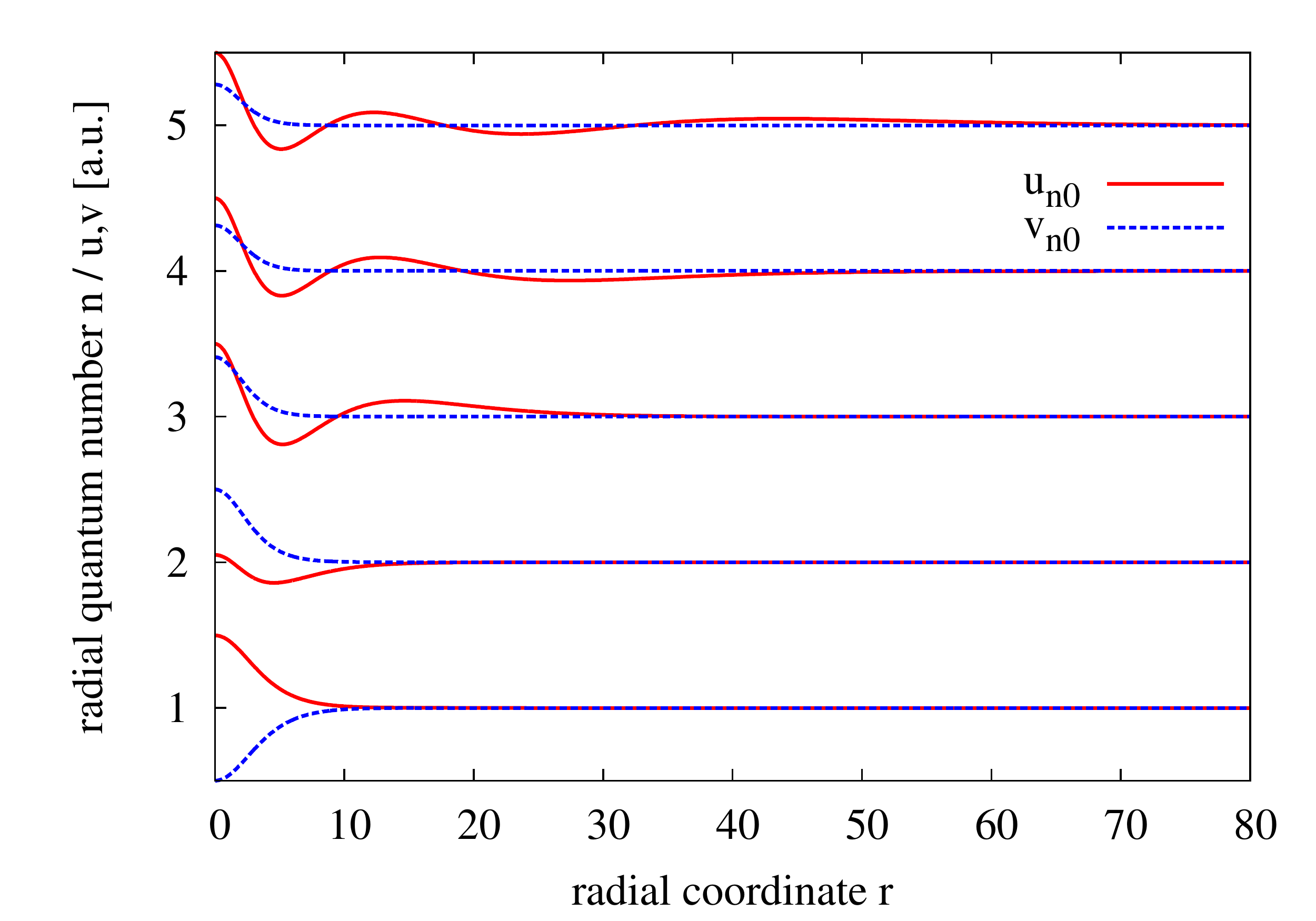}
  \caption{(Color online) Bogoliubov functions $u_{nl}(r)$ and
    $v_{nl}(r)$ for the angular momentum $l=0$ and the radial quantum
    numbers $n=1,\dots,5$. The scattering length is $a=-0.4$. The mode
    with $n=1$ represents the gauge mode discussed in
    Sec.~\ref{sec:bogoliubov-de-gennes}, and the functions $u_{10}(r)$
    and $v_{10}(r)$ are equal to the stationary solution $\psi_0$,
    except for the sign. The functions $u_{n0}(r)$ have $n-1$ nodes,
    whereas all functions $v_{n0}(r)$ show qualitatively the same
    behavior and are nodeless for all $n$. It can be seen that with
    growing radial quantum number the functions $u_{n0}(r)$ extend to
    ever increasing values of $r$.}
  \label{fig:mono_bogo_solutions}
\end{figure}

In Fig.~\ref{fig:mono_bogo_solutions} the Bogoliubov functions $u$ and
$v$ are shown for the angular momentum $l=0$. The lowest functions
with $n=1$ and $n=2$ are concentrated near the origin and have the
same extension as the wave function of the stationary solution (see
Fig.~\ref{fig:mono_wavefunction_a_-0_4_min}). For the higher modes,
the functions $u$ extend further out, which is a consequence of 
the missing external trapping potential.

\subsection{Quantum defect analysis of the Bogoliubov spectrum}
\label{sec:quant-defect-analys}

\begin{figure}
  \includegraphics[width=\columnwidth]{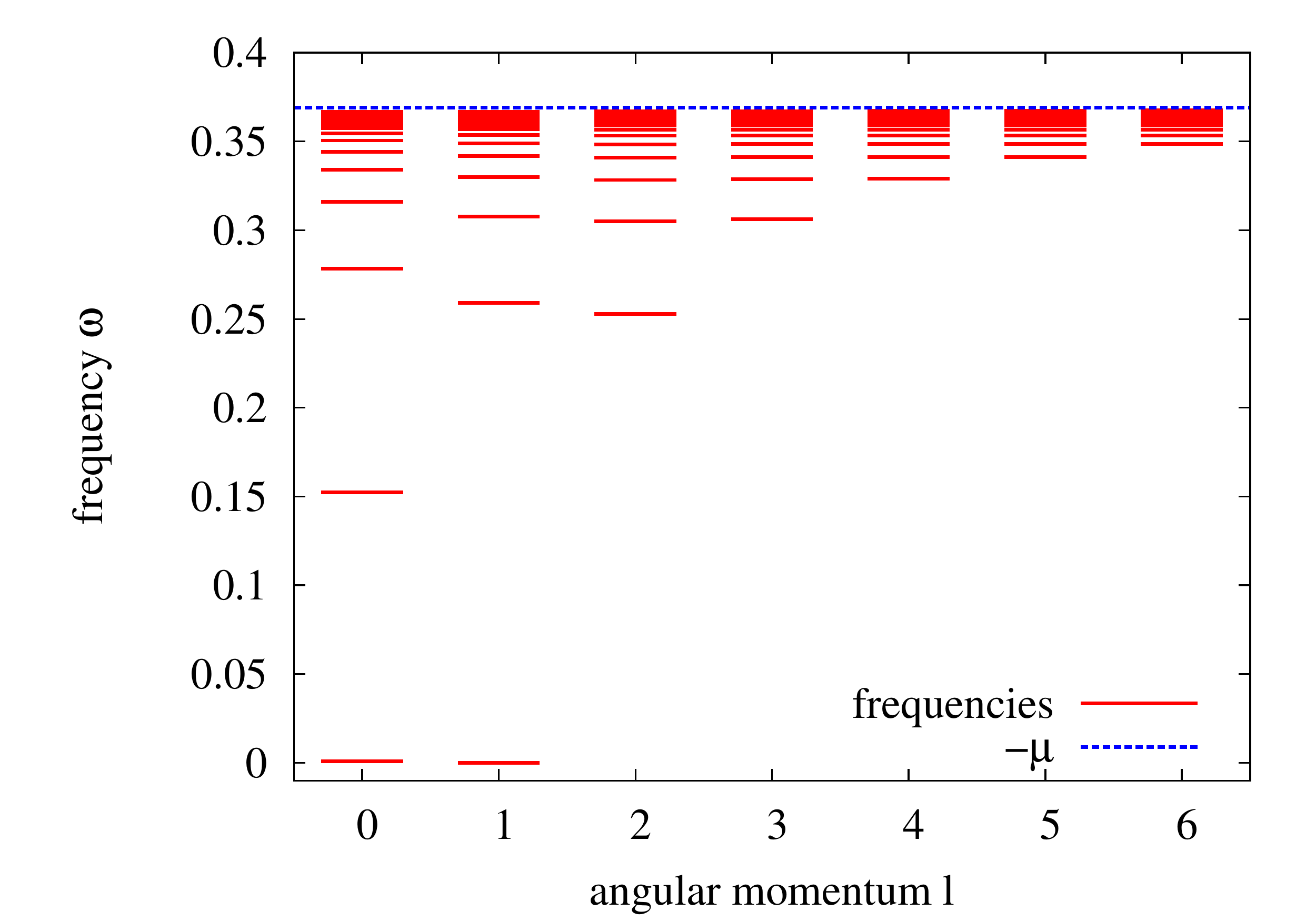}
  \caption{(Color online) Frequencies of the Bogoliubov excitations
    of the ground state of a self-trapped monopolar BEC
    for a fixed scattering length $a=-0.4$, plotted for different values 
    of the
    angular momentum. The dotted line gives the value of the chemical 
    potential. Obviously, as observed in Fig.~\ref{fig:mono_bogo_min} and
    Fig.~\ref{fig:mono_bogo_max}, the frequencies converge to a common limit,
    which is the chemical potential.
    The Bogoliubov spectrum can be described
    by a Rydberg formula with quantum defects.}
  \label{fig:mono_spectrum_a_-0_4_min}
\end{figure}

To prove that for given scattering length the frequencies of the
Bogoliubov excitations converge to a limiting frequency we determined
the $20$ lowest modes for the angular momenta $l=0$ to $6$.  As an
example, Fig.~\ref{fig:mono_spectrum_a_-0_4_min} shows, for the
scattering length $a=-0.4$, the Bogoliubov spectrum of the ground
state.  The convergence of the frequencies to a common limit,
independent of $l$, is evident.  The spectrum is reminiscent of
Rydberg spectra known from alkali atoms. Similar to the spectra of
these atoms, the structure of the Bogoliubov spectra can be understood
in terms of quantum defect theory.

\begin{figure}
  \includegraphics[width=\columnwidth]{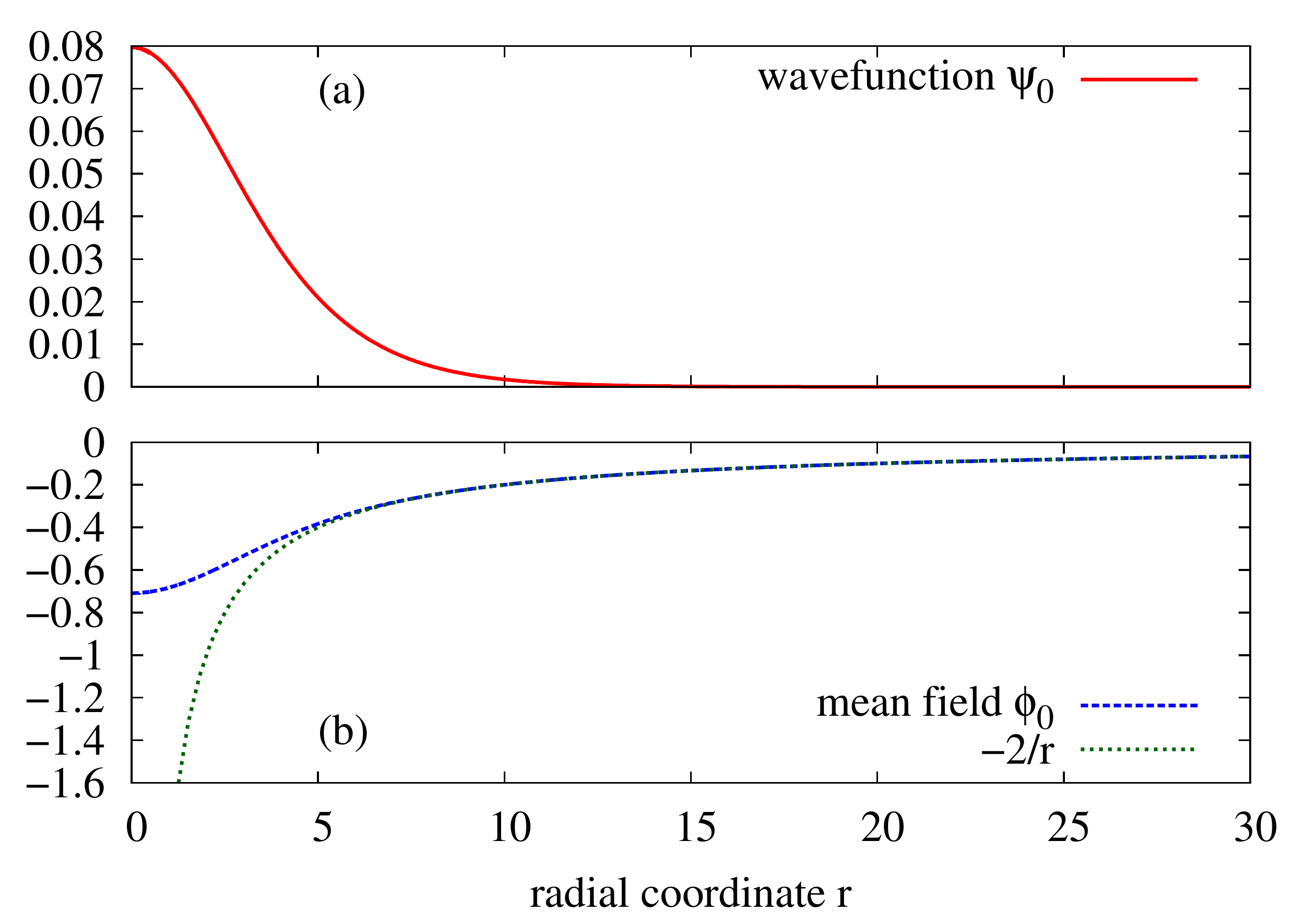}
  \caption{(Color online) (a) Wave function $\psi_0$ and (b)
    mean-field potential $\phi_0$ for the ground state of a
    self-trapped monopolar condensate at a scattering length of
    $a=-0.4$ as functions of the radial coordinate $r$. The wave
    function approaches zero exponentially, whereas the mean-field
    potential behaves like $-2/r$ for large values of $r$. In this
    region, the wave function can be neglected and the mean-field
    potential replaced by its asymptotic form in the BDG equations.}
  \label{fig:mono_wavefunction_a_-0_4_min}
\end{figure}

For large values of $r$ the BDG equations (\ref{eq:bdgmonosph})
simplify due to the fact that the wave function decays exponentially,
and the mean-field potential converges to $\phi_0(r) \approx -2/r$
(see Fig.~\ref{fig:mono_wavefunction_a_-0_4_min}).  Setting
$\psi_0(r) \approx 0$ for $r > r_\text{c}$, all terms containing
$\psi_0$ can be neglected in (\ref{eq:bdgmonosph}), and $\phi_0$ can
be approximated by $-2/r$. This leads to the asymptotic form of the
BDG equations
\begin{subequations}
  \label{eq:bdgmonodefect}
  \begin{align}
    \omega_{nl} u_{nl}(r) &=
      \left[
        - \secondderiv{}{r}
        - \frac{2}{r} \firstderiv{}{r}
        + \frac{l(l+1)}{r^2}
        - \mu
        - \frac{2}{r}
      \right]
      u_{nl}(r), \\
      -\omega_{nl} v_{nl}(r) &=
      \left[
        - \secondderiv{}{r}
        - \frac{2}{r} \firstderiv{}{r}
        + \frac{l(l+1)}{r^2}
        - \mu
        - \frac{2}{r}
      \right]
      v_{nl}(r).
  \end{align}
\end{subequations}
Obviously in this limit $u$ and $v$ obey the same equation, namely the
Schr\"odinger equation of the Coulomb problem, except for the opposite
sign of the eigenvalues. Therefore asymptotically only one equation of
(\ref{eq:bdgmonodefect}) needs to be considered (which will be the one
for $u$). The scattering length enters into Eqs.
(\ref{eq:bdgmonodefect}) only indirectly via $\mu = \mu(a)$.

The approximations made are only valid, if the function values of $u$
and $v$ are small for $r < r_\text{c}$. Especially for lower angular
momenta this is not the case. In the physics of alkali metals a
similar problem occurs: The valence electron far away from the nucleus
``feels'' an attractive $-1/r$ potential, which results from the
shielding of the core electrons. Close to the nucleus, the core
electrons and the true nuclear potential has to be considered.  A
similar situation happens here,
cf. Fig.~\ref{fig:mono_wavefunction_a_-0_4_min}. To account for the
deviation of the potential from the pure Coulomb potential at smaller
values of the radial coordinate we can also introduce a quantum defect
in the formula for the Rydberg series eigenvalues (see, e.g.,
\cite{Seaton83a}),
\begin{align}
  \label{eq:bdgmonodef}
  \omega_{nl} = -\mu - \frac{1}{(n+l+1-\delta_l)^2},
\end{align}
where the quantum defects $\delta_l$ depend on the angular
momentum. The negative chemical potential is the limit of the
frequencies for $n \to \infty$. The quantum defects can be obtained by
least-squares fits of the Bogoliubov frequencies $\omega_{nl}$ to
Eq.~(\ref{eq:bdgmonodef}). They converge to constant values for large
$n$. Since Eq.~(\ref{eq:bdgmonodef}) strictly holds only in this
limit, in the fits it can be necessary to neglect the lowest
frequencies.

For growing angular momentum, the repulsive effective potential
$l(l+1)/r^2$ becomes stronger, and this centrifugal barrier ensures
that the absolute values of the functions $u$ and $v$ decrease close
to the origin $r = 0$. This leads to a smaller quantum defect
$\delta_l$, since the approximation made in deriving
Eqs.~(\ref{eq:bdgmonodefect}) becomes valid at smaller values of
$r$. In accordance with the quantum defects in alkalis
\cite{Seaton83a}, the quantum defects therefore will tend to zero for
higher angular momenta.

\begin{figure}
  \includegraphics[width=\columnwidth]{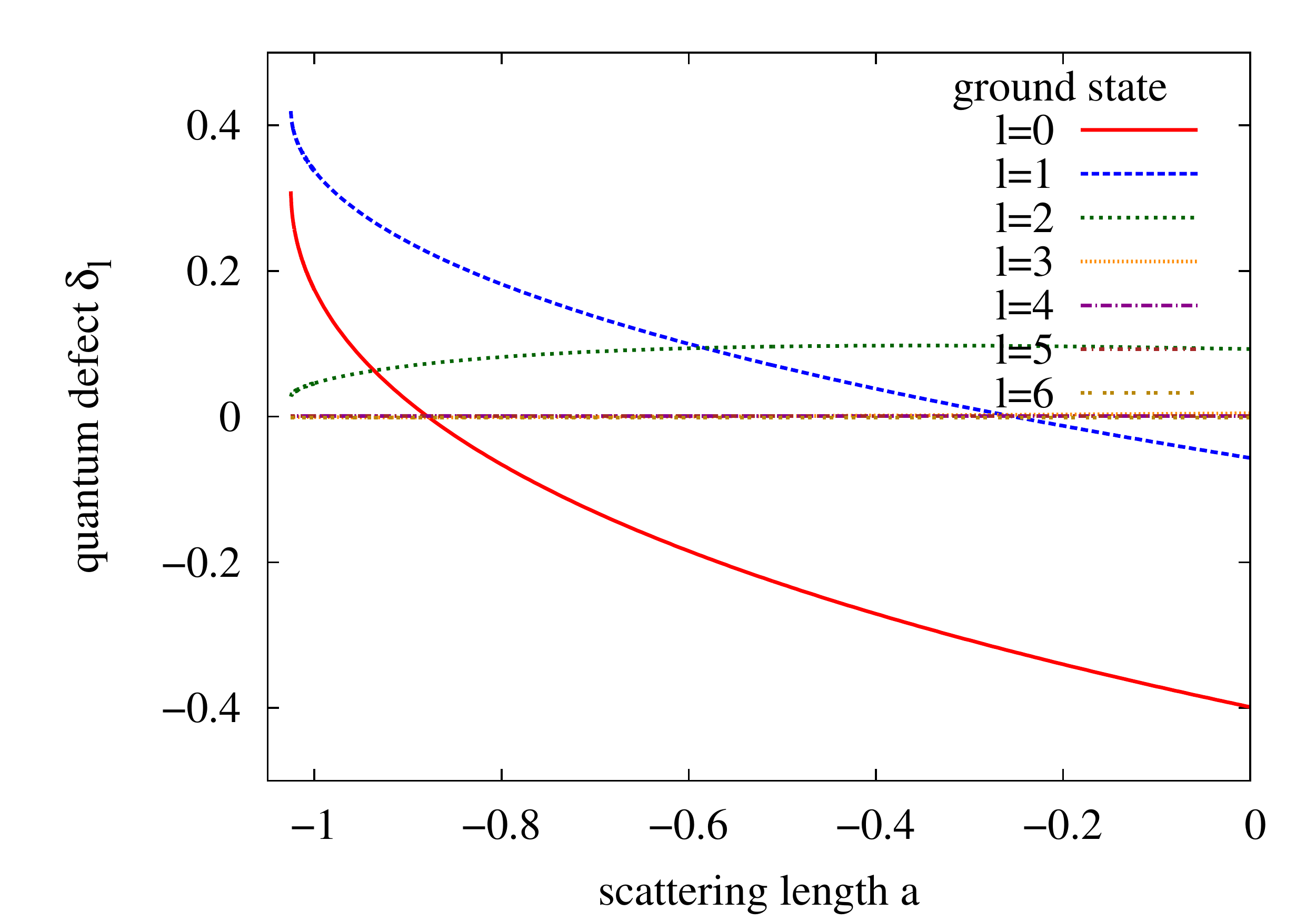}
  \caption{(Color online) Calculated quantum defects $\delta_l$ for
    the Bogoliubov spectrum of the ground state
    (Fig.~\ref{fig:mono_bogo_min}) in dependence of the scattering
    length $a$ for different angular momenta $l$.  The quantum defects
    for $l=0$ and $l=1$ rise steeply and turn from negative to
    positive as the scattering length is decreased, while the quantum
    defect for $l=2$ shows only a weak dependence on the scattering
    length and drops close to the critical scattering length. As
    expected, for the higher angular momenta $l > 2$ the quantum
    defects are close to zero.}
  \label{fig:mono_quantum_defects_min}
\end{figure}

\begin{figure}
  \includegraphics[width=\columnwidth]{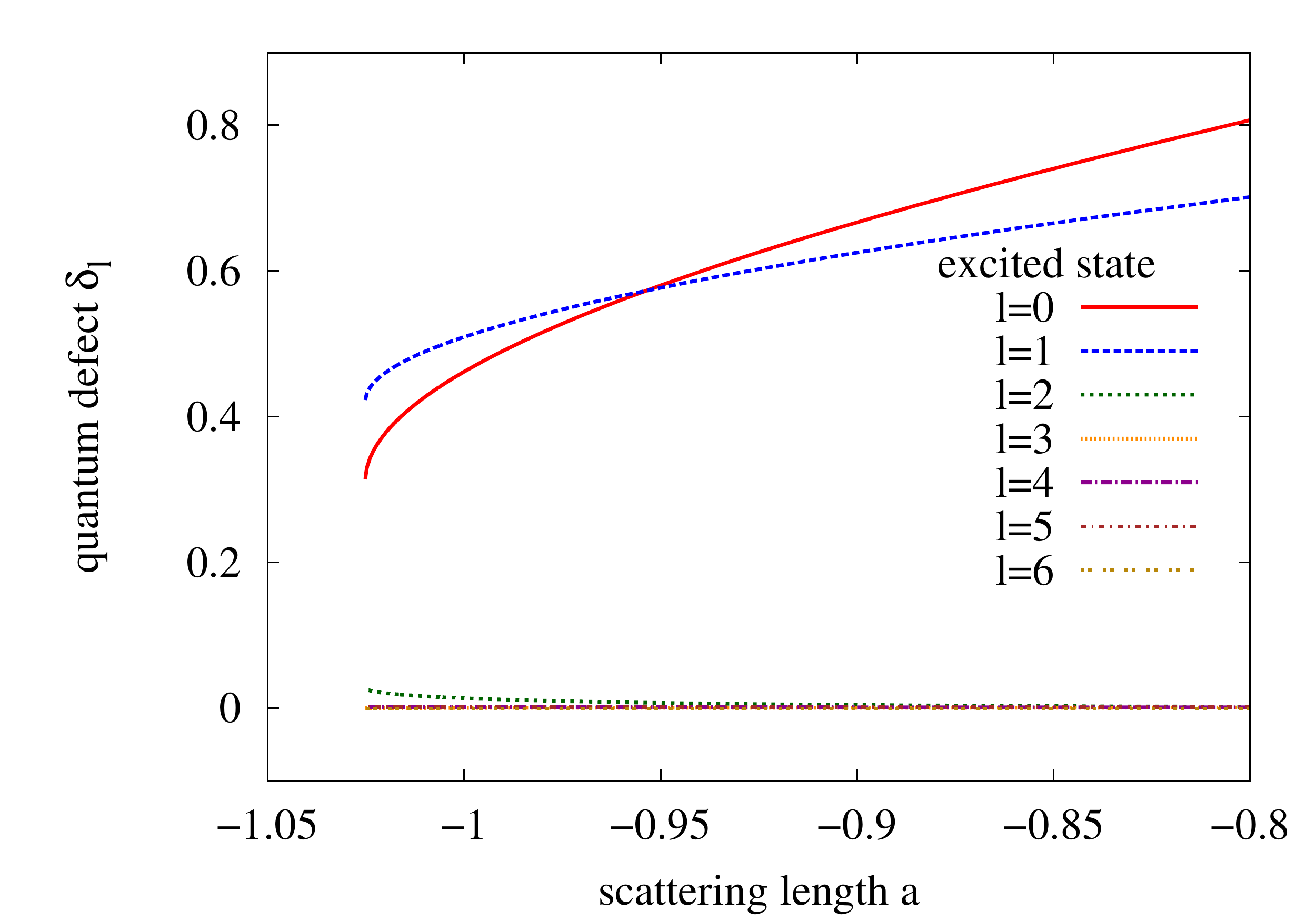}
  \caption{(Color online) Same as
    Fig.~\ref{fig:mono_quantum_defects_min}, but for the excited
    state. Since the Bogoliubov spectra of the ground and excited
    state merge at the critical scattering length, the same holds for
    the quantum defects. The quantum defects $\delta_0$ and $\delta_1$
    grow as the scattering length is increased, whereas $\delta_2$
    drops and tends to zero. As in the case of the ground state, the
    quantum defects for $l > 2$ are close to zero.}
  \label{fig:mono_quantum_defects_max}
\end{figure}

In Fig.~\ref{fig:mono_quantum_defects_min} we present the quantum
defects calculated for the Bogoliubov excitations of the ground
state. Obviously the quantum defects for $l=0$ and $l=1$ show a strong
dependence on the scattering length, while for $l\ge 2$ they are
almost constant, and in particular close to zero for $l>2$.
Eq.~(\ref{eq:bdgmonodef}) reproduces the frequencies of the Bogoliubov
excitations of the ground state for all modes with an absolute error
of less than $10^{-3}$, except for the two lowest $l=0$ modes and the lowest
$l=1$ mode. The quantum defect analysis for the Bogoliubov excitations
of the excited state is presented in
Fig.~\ref{fig:mono_quantum_defects_max}. The quantitative statements
made for the excitations of the ground state also hold for this state.
The only difference is that the quantum defect for $l=2$ tends to zero
as the scattering length is increased.

Thus by means of quantum defect analysis we have been able to explain
the Rydberg-like structure of the Bogoliubov spectra of the ground and
excited state of self-trapped monopolar BECs, 
and could confirm that the negative chemical potential
is the limit of the frequencies for all angular momenta.

\section{Variational approach with Gaussian functions and spherical
  harmonics}
\label{sec:vari-appr-with}

We now turn our attention to variational calculations.
The simplest ansatz with a single Gaussian centered at
the origin was used by Perez-Garcia et al.  \cite{Perez-Garcia96a} to
determine monopolar and quadrupolar modes of BECs without long-range
interactions. The ansatz was improved by using coupled Gaussians
\cite{Heller76a,Heller81a}, and it was shown \cite{Rau10a,Rau10b} that
this method is capable of reproducing accurately the stationary states
even of BECs with long-range interactions, calculated numerically. 
The ansatz employed to
determine the stationary solution of a radially symmetric condensate
was
\begin{align}
  \label{eq:ansatzsph}
  \psi = \sum\limits_{k=1}^N \eto{-A_r^k r^2 -\gamma^k},
\end{align}
where the complex quantities $A_r^k$ and $\gamma^k$ are the widths and
the amplitudes, respectively, of each Gaussian. 
The above ansatz can only describe monopolar excitation modes, 
since the wave function
$\psi$ is independent of the angular coordinates $\theta$ and
$\phi$. If one chooses the widths differently for each space
direction,
\begin{align}
  \label{eq:ansatzquad}
    \psi = \sum\limits_{k=1}^N \eto{-A_x^k x^2 - A_y^k y^2 - A_z^k z^2 -\gamma^k},
\end{align}
the width of a condensate can oscillate independently in each
direction, which represents quadrupolar oscillations.

A generalization of Eqs.~(\ref{eq:ansatzsph})
and~(\ref{eq:ansatzquad}), which includes general square and linear
terms in the exponentials, is \cite{Heller76a,Heller81a}
\begin{align}
  \label{eq:vargaussian}
  \psi =
  \sum\limits_{k=1}^N g^k \equiv
  \sum\limits_{k=1}^N
  \exp
  \left(
    - \vec{r}^\text{T} \mathbf{A}^k \vec{r}
    - (\vec{p}^k)^\text{T} \vec{r}
    - \gamma^k
  \right)\,,
\end{align}
with complex symmetric matrices $\mathbf{A}^k$, complex vectors
$\vec{p}^k$ and complex numbers $\gamma^k$. This ansatz can describe
excitation modes with angular momenta up to $l=2$.  To see this
consider a small deviation $\delta\vec{z}$ of the variational
parameters from those of a stationary solution $\vec{z}_0$ and Taylor
expand the ansatz with coupled Gaussians~(\ref{eq:vargaussian}) for the
perturbed wave function $\psi(\vec{z}_0 + \delta\vec{z})$ to first
order in $\delta\vec{z}$,
\begin{align}
  \delta\psi &= \delta\vec{z} \cdot
  \left.
    \firstpderiv{\psi}{\vec{z}}
  \right|_{\vec{z}=\vec{z}_0} \nonumber \\
  &= -\sum\limits_{k=1}^N
  \left(
    \vec{r}^\text{T} \delta\mathbf{A}^k \vec{r}
    + (\delta\vec{p}^k)^\text{T} \vec{r}
    + \delta\gamma^k
  \right)
  \left.
    g^k
  \right|_{\vec{z}=\vec{z}_0}.
\end{align}
Since only terms at most quadratic in $x,y,z$ appear in front of the
exponentials, these terms can be expressed by spherical harmonics with
angular momenta $l=0,1,2$, which proves our statement.

We apply an ansatz which is capable of 
describing excitations with -- in principle -- arbitrary angular momenta. 
Motivated by the separation in
the BDG equations with spherical harmonics in
Eq.~(\ref{eq:bdgmonosep}), we directly include the spherical harmonics
in an extended variational ansatz
\begin{align}
  \label{eq:varansatzsph}
  \psi = \sum\limits_{k=1}^N
  \left(
    1 +
    \sum\limits_{(l,m) \neq (0,0)} d_{lm}^k Y_{lm} (\theta,\phi) r^l
  \right)
  \eto{-A_r^k r^2 - \gamma^k}.
\end{align}
The amplitudes $d_{lm}^k$ account for additional angular momenta
$(l,m)$. The sum over $(l,m)$ may include arbitrary angular momenta,
adjusted to the problem. For instance, if one wishes to calculate the
linear perturbation of a specific angular momentum $(l,m)$, as we do
below, the sum in Eq.~(\ref{eq:varansatzsph}) needs to include $(l,m)$
and $(l,-m)$, since the nonlinear terms in the GPE lead to a coupling
of different angular momenta.

\subsection{Equations of motion and stability analysis}
\label{sec:equations-motion}

In order to carry out calculations with the extended variational
ansatz (\ref{eq:varansatzsph}), we need the equations of motion for
the variational parameters. We use the approach of \cite{Rau10a} based
on the Dirac-Frankel-McLachlan time-dependent variational principle
\cite{McLachlan64a,Dirac30a}. An arbitrary ansatz for the wave
function is made $\psi = \psi (\vec{z})$, with the -- in general
complex -- variational parameters $\vec{z} = (z_1,\dots,z_M)$, for a
system governed by the Schr\"odinger equation
\begin{align}
  \label{eq:schrod}
  \ii \dot{\psi} = \hat{H} \psi,
\end{align}
where the Hamiltonian $\hat{H}$ may contain nonlinear terms in the
wave function.  The principle states that the norm of the difference
between the left- and the right-hand side of (\ref{eq:schrod})
\begin{align}
  I = ||\ii \phi(t) - \hat{H} \psi(t)||^2
\end{align}
must be minimized. For a fixed time $t$, $\psi(t)$ is given, and $I$
is minimized by varying $\phi(t)$. After the minimization, $\phi$ is
set to $\phi = \dot{\psi}$. A necessary condition for the minimization
of $I$ is \cite{Cartarius08b}
\begin{align}
  \label{eq:tdvpeom}
  \mathbf{K} \dot{\vec{z}} = -\ii \vec{h},
\end{align}
the matrix $\mathbf{K}$ and the vector $\vec{h}$ are defined by
\begin{subequations}
  \begin{align}
    K_{ij} &=
    \qmprod{\firstpderiv{\psi}{z_i}}{\firstpderiv{\psi}{z_j}}, \\
    h_i &= \qmprod{\firstpderiv{\psi}{z_i}}{\hat{H} \psi}.
  \end{align}
\end{subequations}
Stationary solutions can then be found by requiring
\begin{align}
  \dot{z}_i = -\ii \sum\limits_{j=1}^M (\mathbf{K}^{-1})_{ij} h_j =
  \begin{cases}
    \ii \mu & \text{for } z_i \equiv \gamma^k, \\
    0 & \text{else},
  \end{cases}
\end{align}
and searching for $\vec{z}$ in a nonlinear root search.

The stability properties and linear oscillations of a stationary
solution $\vec{z}_0$ can be found by first changing from the complex
$M$-dimensional vector $\vec{z}$ to a real $2M$-dimensional vector
$\tilde{\vec{z}}$ containing the real and imaginary parts of the
variational parameters, and considering a small perturbation,
$\tilde{\vec{z}}(t) = \tilde{\vec{z}}_0 +
\delta\tilde{\vec{z}}(t)$. Linearization of the equations of motion
(\ref{eq:tdvpeom}) yields the time dependency of the perturbation
\cite{Rau10a}
\begin{align}
  \delta\dot{\tilde{\vec{z}}}(t) = \mathbf{J} \delta\tilde{\vec{z}}(t)
\end{align}
with the Jacobian
\begin{align}
  J_{ij} = \firstpderiv{\dot{\tilde{z}}_i}{\tilde{z}_j}
\end{align}
evaluated at the fixed point $\tilde{\vec{z}} =
\tilde{\vec{z}}_0$. The excitation modes of the stationary solutions
are finally found by diagonalizing the Jacobian $\mathbf{J}$.

All integrals appearing in Eq.~(\ref{eq:tdvpeom}) with the ansatz
(\ref{eq:varansatzsph}) can be calculated analytically. The contact
interaction leads to integrals over four spherical harmonics which can
be expressed in terms of Wigner-3j symbols.  The contribution of
long-range monopolar potential can be evaluated by inserting the
multipole expansion for the monopolar integration kernel, which leads
to Gaussian integrals. For further details of the calculations we
refer to the appendix.

\subsection{Test in a system without long-range interactions}
\label{sec:test-system-without}

\begin{figure}
  \includegraphics[width=\columnwidth]{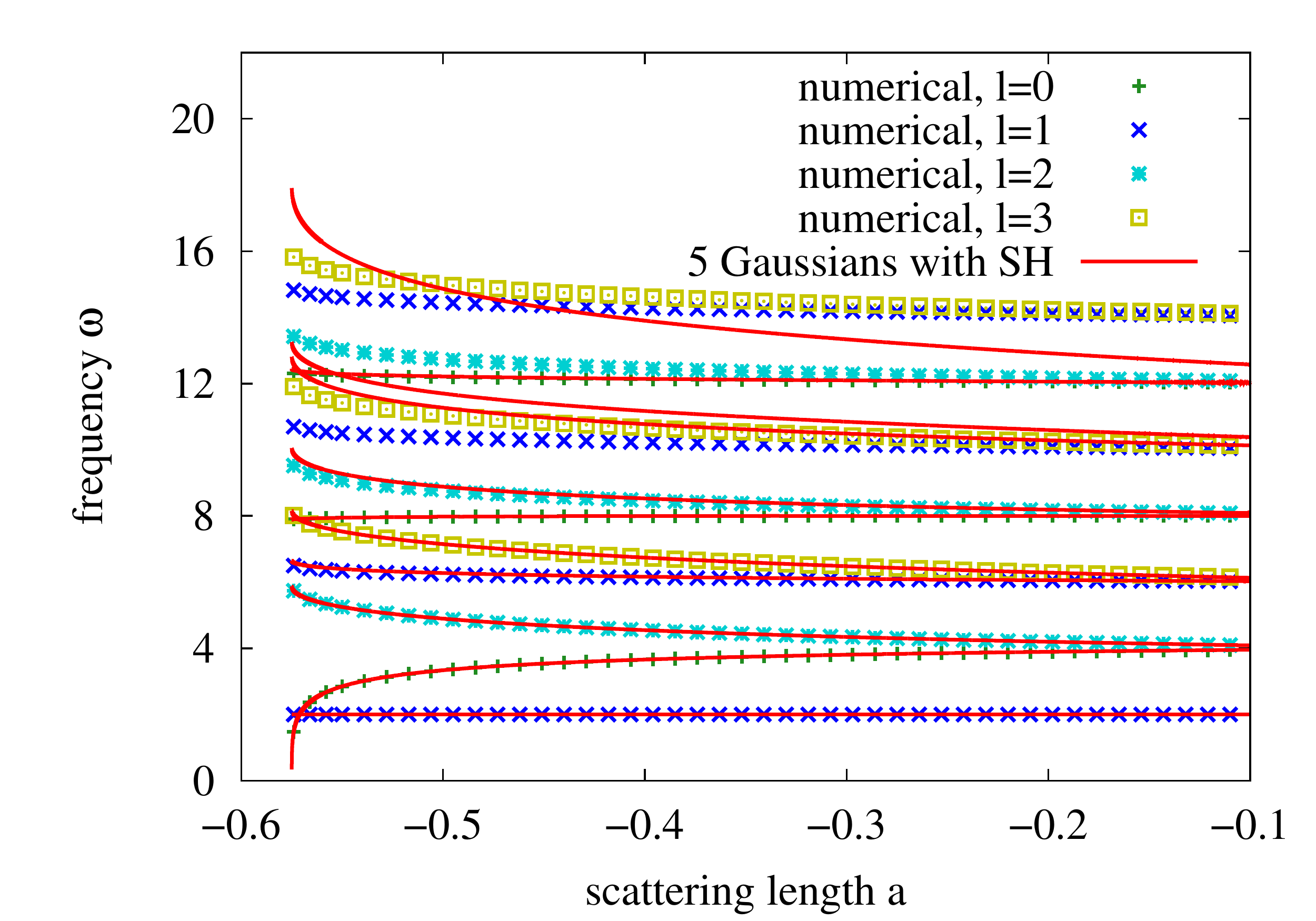}
  \caption{(Color online) Comparison of the full-numerical Bogoliubov
    spectrum of a BEC with attractive contact interaction with the
    spectrum obtained from the variational ansatz with coupled
    Gaussians and spherical harmonics (SH). The variational ansatz has
    been used with $5$ coupled Gaussians and spherical harmonics up to
    an angular momentum of $l=3$. For the lowest modes we find
    excellent agreement. There are almost no deviations for
    frequencies $\omega < 10$. Just slightly above the critical
    scattering length small differences can be seen in the figure. For
    the higher modes, differences become larger and the variational
    ansatz can describe the Bogoliubov modes only qualitatively
    correct.}
  \label{fig:wo_lr_int_5gauss_sph_harm_bogo_min}
\end{figure}

As a first test we apply the extended variational ansatz
(\ref{eq:varansatzsph}) to a condensate in a radially symmetric trap
with short-range scattering interaction. The GPE for this system reads
\begin{align}
  \label{eq:gpetimeshort}
  \ii \firstpderiv{\psi}{t} (\vec{r},t) = 
  \left[
    - \Delta
    + r^2
    + 8 \pi a \abs{\psi(\vec{r},t)}^2
  \right]
  \psi(\vec{r},t).
\end{align}
Here units based on the trapping frequency $\gamma = \omega/2$ and the
harmonic oscillator length $a_0 = \sqrt{\hbar/m \omega}$ have been
used. The scaled dimensionless scattering length $a$ in
(\ref{eq:gpetimeshort}) corresponds to $N a / a_0$ in SI units, with
the particle number $N$. These units will be used in all figures for
the condensate without long-range interaction. The BDG equations are
given in Eqs.~(\ref{eq:bdgmono}) and~(\ref{eq:bdgmonosph}),
respectively, with all terms containing the mean-field potential
$\phi_0$ and the auxiliary field $f$ omitted, and the trapping
potential $V_\text{ext} = r^2$ included.

The BDG equations for condensates with short-range interaction were
first solved numerically by \cite{Edwards96a,Ruprecht96a}. In this work, we used
the method discussed in Sec.~\ref{sec:bogoliubov-de-gennes}.

\begin{figure}
  \includegraphics[width=\columnwidth]{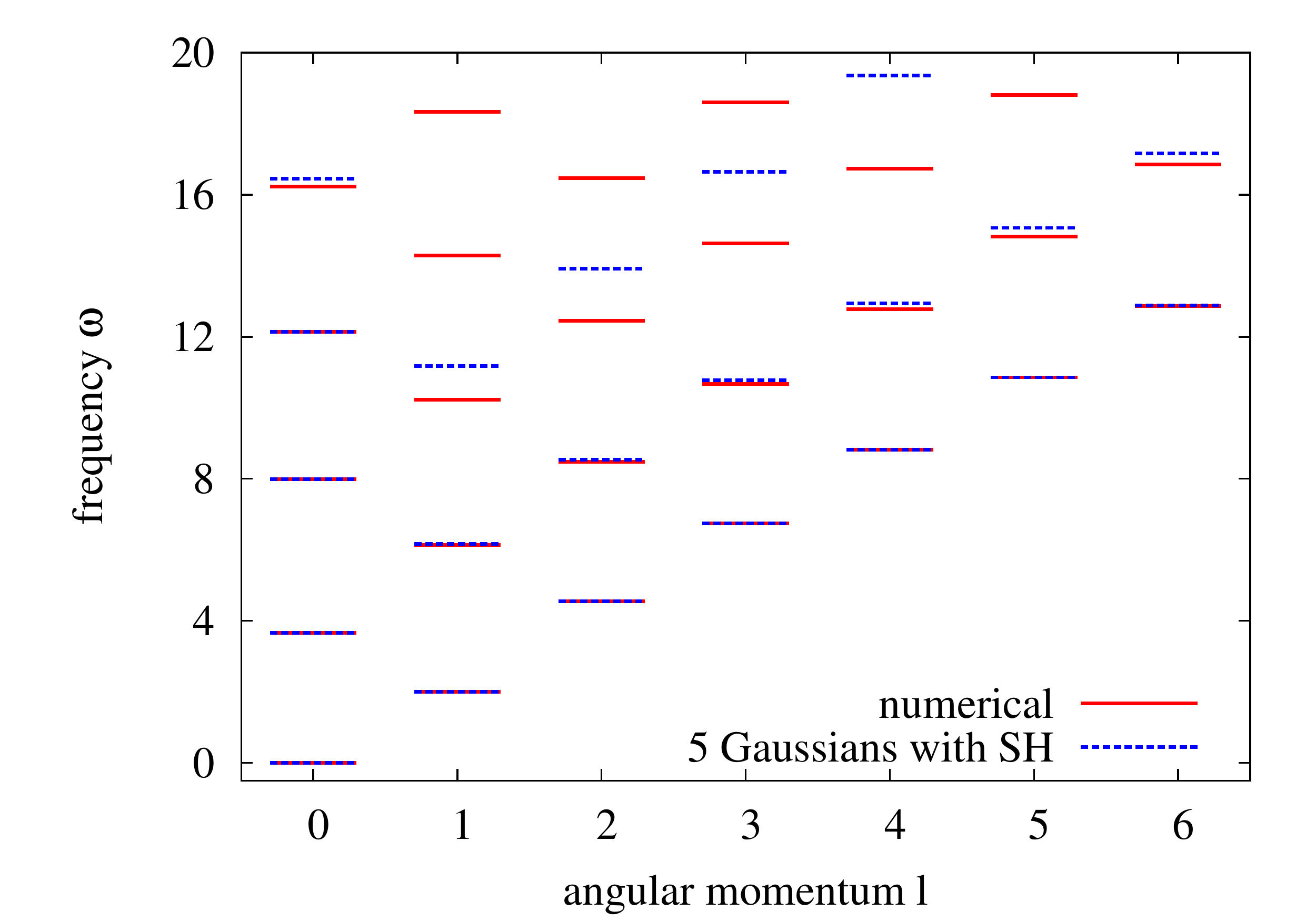}
  \caption{(Color online) Comparison of both spectra as in
    Fig.~\ref{fig:wo_lr_int_5gauss_sph_harm_bogo_min}, but here for a
    fixed scattering length of $a=-0.4$ and angular momenta up to
    $l=6$.  For $l=0$ the variational ansatz reproduces the Bogoliubov
 frequencies very well for the four lowest  modes, and  with only small 
deviations for the two lowest modes in the 
    higher angular momentum bands. }
  \label{fig:wo_lr_int_spectrum_a_-0_4_min}
\end{figure}

Fig.~\ref{fig:wo_lr_int_5gauss_sph_harm_bogo_min} shows the
eigenfrequencies of the Bogoliubov excitations of the ground state
with $l= 0, 1, 2$ and $3$ as functions of the scattering length. For
$a=0$ one obtains the equidistant eigenfrequencies of the harmonic
oscillator. When the scattering length is decreased the attractive
short-range interaction acts as a perturbation, and the frequencies
are slightly shifted. For $a \to a_\text{crit} \approx -0.0575$ the
lowest $l=0$ mode drops to zero marking the collapse of the
condensate. The lowest mode with $l=1$ represents the oscillation of
the center-of-mass of the condensate, and its value is exactly that of
the trapping frequency $\omega = 2\gamma = 2$.

For comparison in Fig.~\ref{fig:wo_lr_int_5gauss_sph_harm_bogo_min} we
also show the results for the eigenvalues of the Jacobian matrix at the
ground state fixed point obtained in the variational ansatz 
using 5 Gaussians in
combination with spherical harmonics (\ref{eq:varansatzsph}). One
recognizes that in particular the eigenvalues of the lowest modes in
each angular momentum band excellently agree with the eigenfrequencies
of the Bogoliubov excitations. It is only close to the critical
scattering length that small deviations appear. The lowest
center-of-mass excitation with $l=1$ can even be reproduced within
numerical accuracy, independent of the number of Gaussians used. For
the higher modes with eigenvalues of the Jacobian $\omega > 10$, only far away
from the critical point the variational and full-numerical results
still approximately correspond to each other, and in the vicinity of the
critical scattering length the Jacobi eigenvalues can reproduce the
behavior of the Bogoliubov excitation eigenfrequencies only qualitatively.

We also tested the variational ansatz (\ref{eq:varansatzsph}) for
higher angular momenta up to $l=6$. The results for a fixed scattering
length of $a=-0.4$ are presented in
Fig.~\ref{fig:wo_lr_int_spectrum_a_-0_4_min}. One recognizes a very
good agreement for the lowest modes in each $l$ band, and small
differences for the second-lowest modes. This demonstrates that for
condensates with attractive short-range interaction the eigenvalues of
the Jacobian matrix calculated at the fixed point corresponding to the
ground state in the new variational ansatz indeed quantitatively
coincide with the eigenfrequencies of the lowest Bogoliubov modes.

\subsection{Application of the variational approach to the monopolar
  condensate}
\label{sec:appl-monop-cond}

\begin{figure*}
  \includegraphics[width=0.44\textwidth]{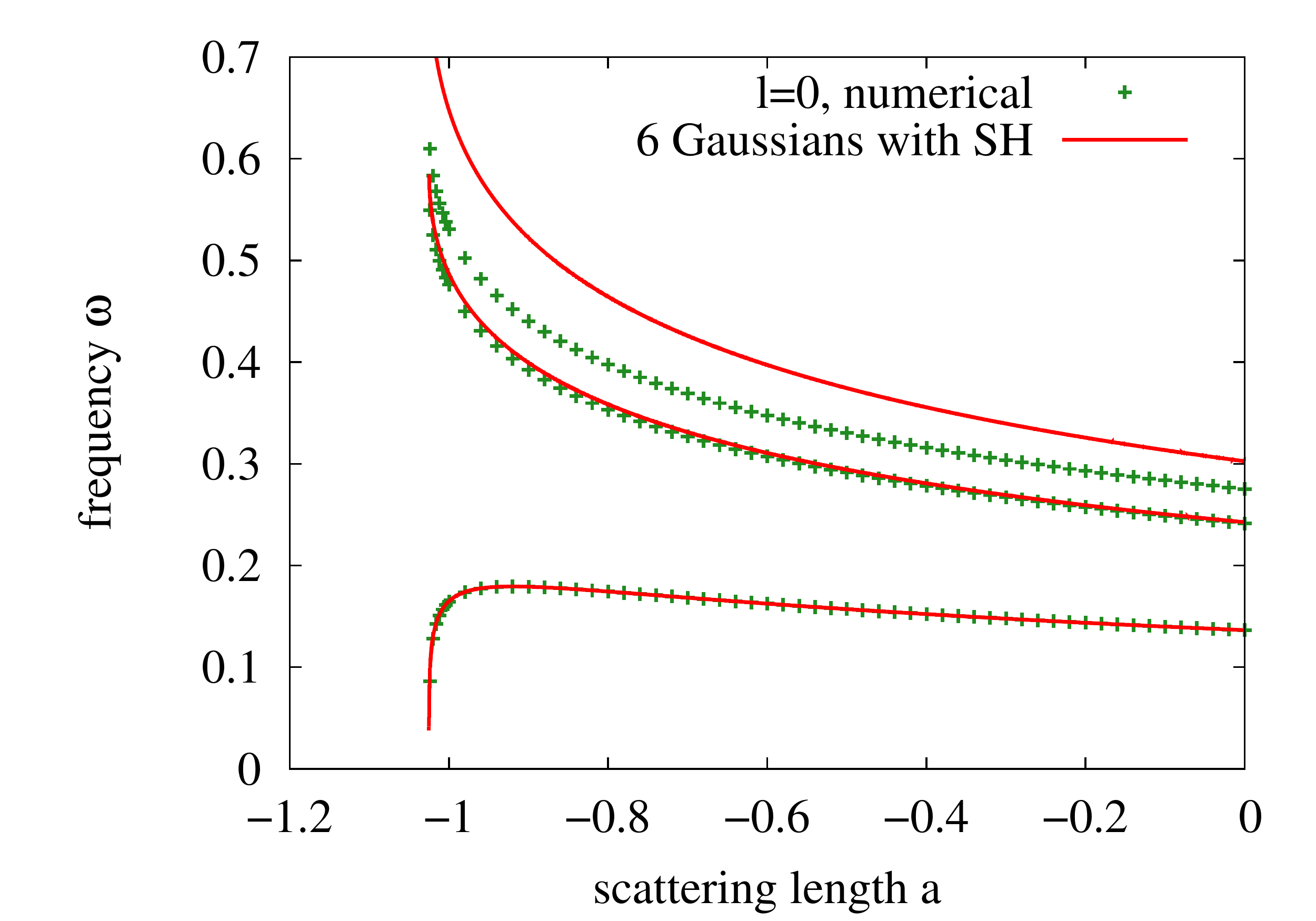}
  \includegraphics[width=0.44\textwidth]{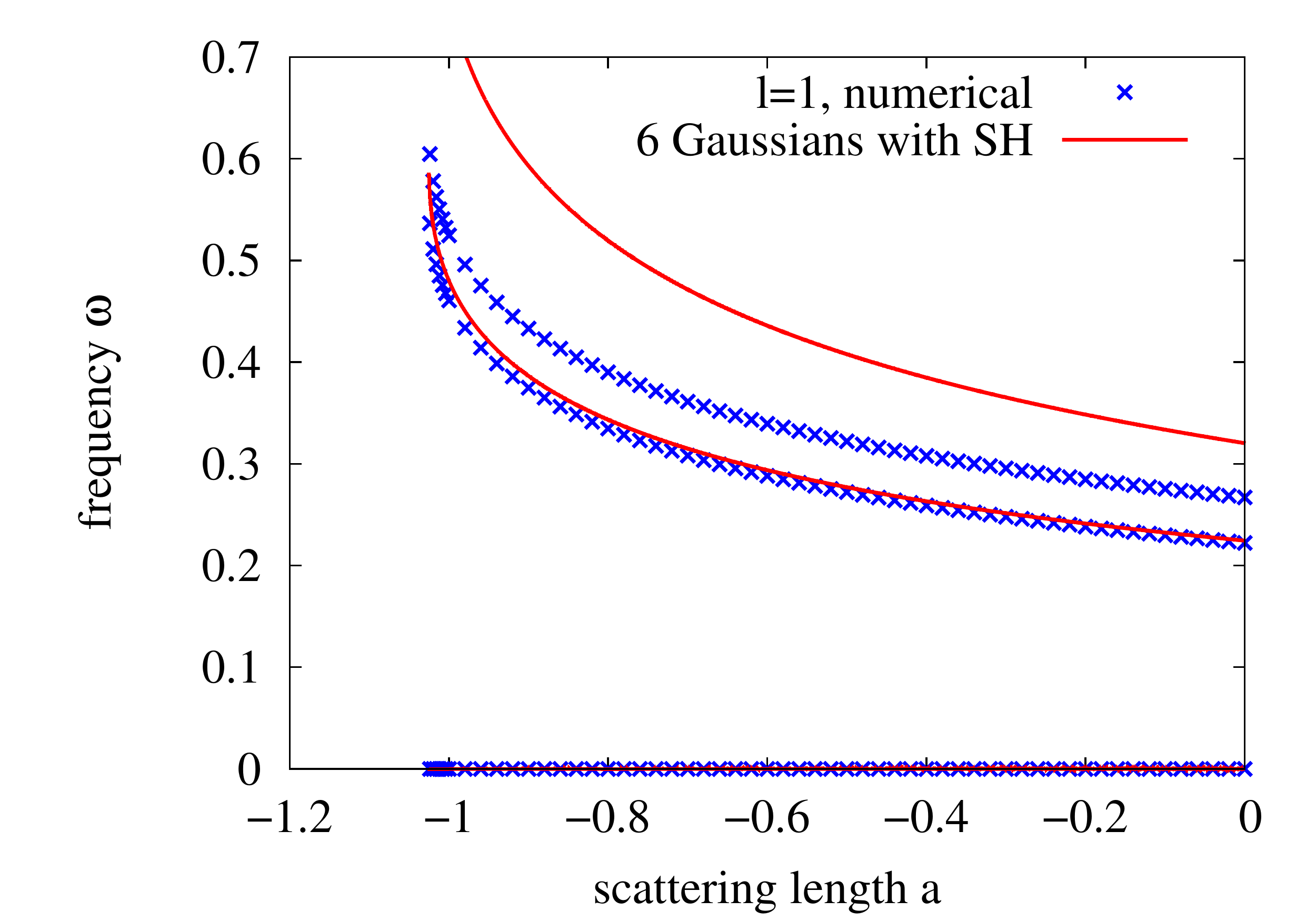}
  \includegraphics[width=0.44\textwidth]{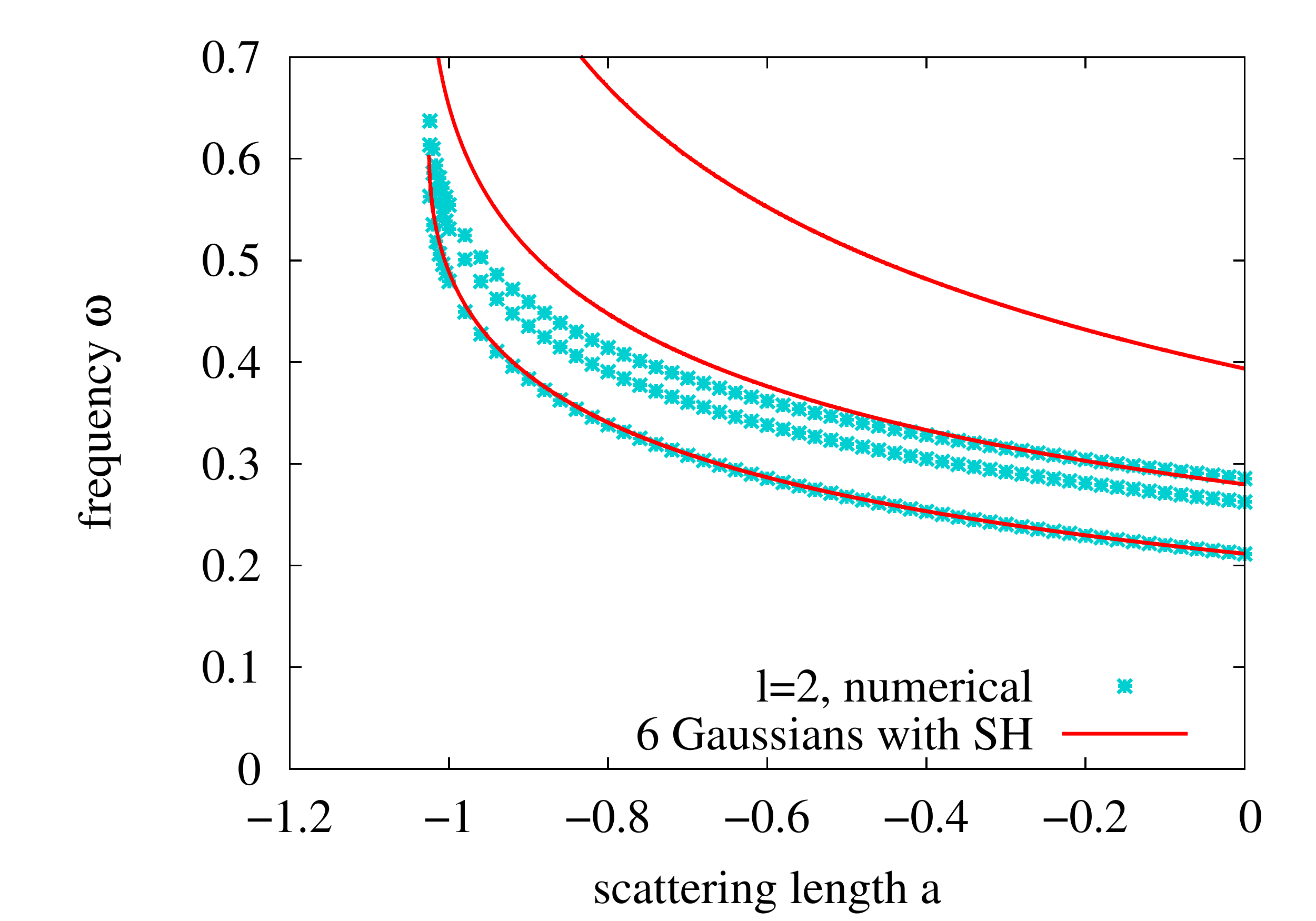}
  \includegraphics[width=0.44\textwidth]{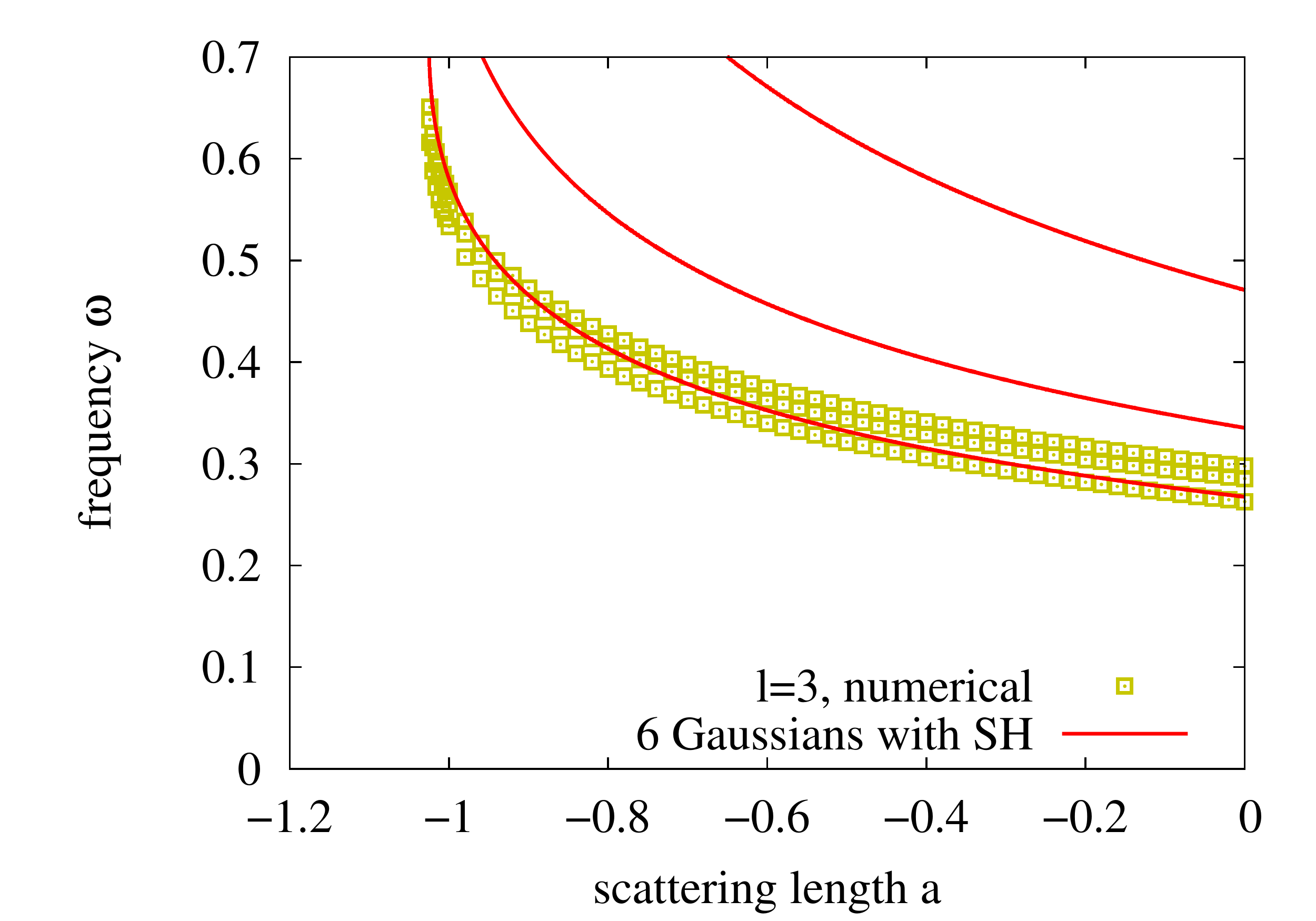}
  \caption{(Color online) Comparison of the full-numerical Bogoliubov
    spectrum of the ground state of a self-trapped monopolar BEC with
    the spectrum obtained from the variational ansatz with coupled
    Gaussians and spherical harmonics (SH). The variational ansatz has
    been used with $6$ coupled Gaussians and spherical harmonics up to
    an angular momentum of $l=3$. For the lowest $l=0$ and $l=1$ mode
    the results of both methods almost cannot be distinguished.  The
    differences of the frequencies of the second lowest $l=0$ and
    $l=1$ and the lowest $l=2$ modes are small in the range of the
    scattering length considered. The lowest $l=3$ mode is well
    approximated by the variational ansatz, but the differences in
    frequency are larger compared to the frequencies belonging to
    lower angular momenta. For the higher modes there is no
    quantitative agreement.}
  \label{fig:mono_l0_3_6gauss_bogo}
\end{figure*}

We now apply the extended variational ansatz (\ref{eq:varansatzsph})
to the self-trapped monopolar condensate. For the three lowest
excitations Fig.~\ref{fig:mono_l0_3_6gauss_bogo} shows the comparison
of the full-numerical Bogoliubov spectrum with the spectrum obtained
from the eigenvalues of the Jacobian matrix in the variational
ansatz. We used $N=6$ Gaussians and angular momenta up to $l=3$. The
lowest modes for $l=0$ and $l=1$ match very well in the whole range
of scattering lengths considered. For the second-lowest $l=0$ and
$l=1$ and the lowest $l=2$ mode we find a good agreement, but the
differences become larger as the scattering length approaches the
critical point. Nevertheless, we have the result that the variational
ansatz with coupled Gaussians and spherical harmonics is a valid
alternative to the full-numerical quantum mechanical approach
also in this case, if one is interested in these modes.

Looking at the lowest mode with $l=3$ one finds that the agreement
is good for scattering lengths around $a=0$, but the two frequencies
deviate as the scattering length is decreased. The eigenmode of the
variational ansatz can only be seen as an approximation to the
full-numerical one. The other modes can only be described
qualitatively by the variational approach.

\begin{figure}
  \includegraphics[width=0.95\columnwidth]{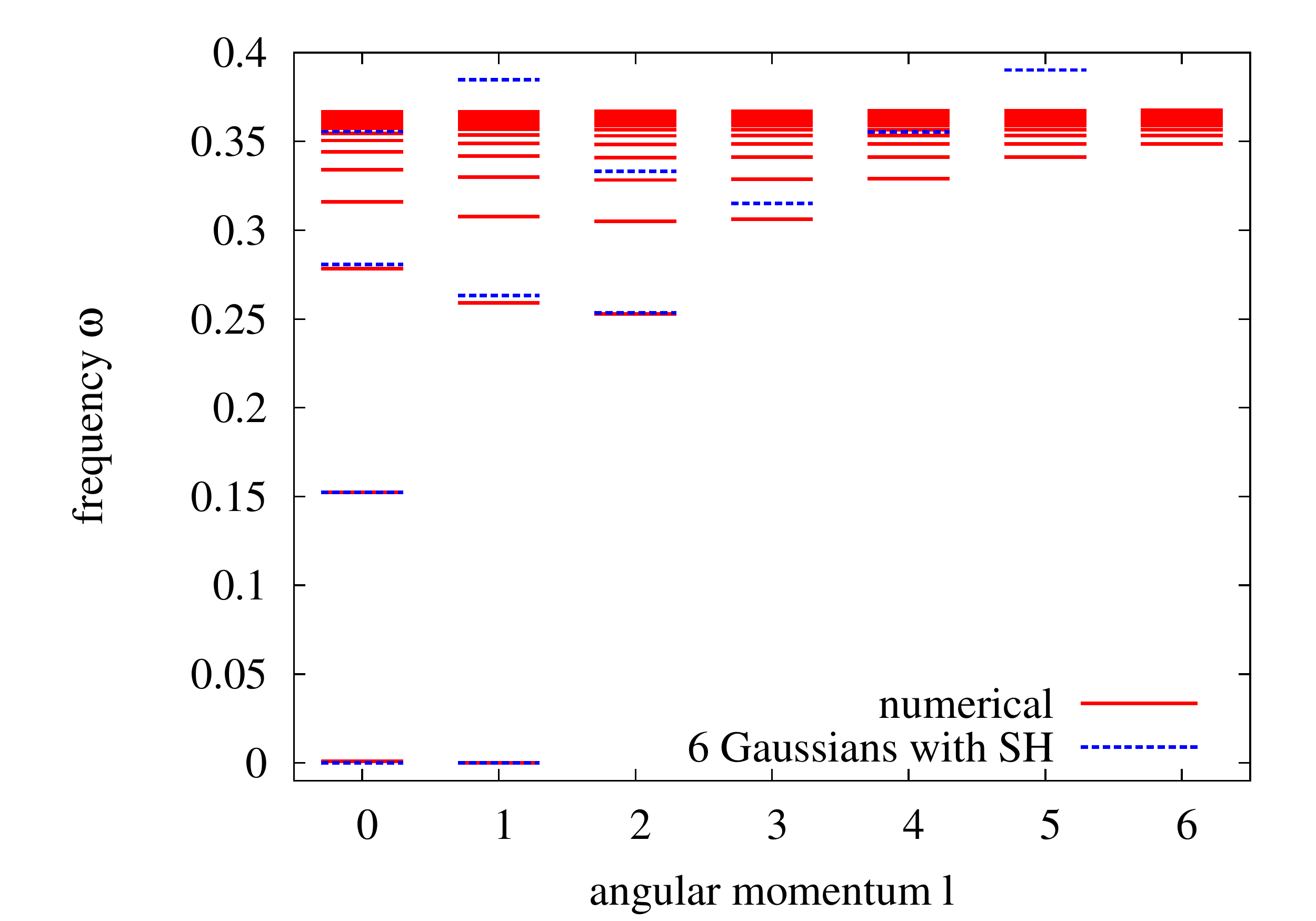}
  \caption{(Color online) Comparison of both spectra as in
    Fig.~\ref{fig:mono_l0_3_6gauss_bogo}, for a self-trapped monopolar
    BEC, at the fixed scattering length of $a=-0.4$ and angular
    momenta up to $l=6$. For $l=0$ and $l=1$ the two lowest modes, and
    for $l=2$ and $l=3$ only the lowest modes, agree well. For higher
    angular momenta $l \geq 5$, the lowest mode lies even above the
    limit of the numerical Bogoliubov spectrum (compare with
    Fig.~\ref{fig:mono_spectrum_a_-0_4_min}).}
  \label{fig:mono_spectrum_gauss_a_-0_4_min}
\end{figure}

We also applied the variational ansatz (\ref{eq:varansatzsph}) for
higher angular momenta up to $l=6$. The results for a fixed scattering
length of $a=-0.4$ are presented in
Fig.~\ref{fig:mono_spectrum_gauss_a_-0_4_min}. As already noticed,
only the lowest modes and angular momenta agree well with the
numerically exact values. In the remaining modes the excitation
frequencies differ distinctly.  For $l=5$, the frequency of the
lowest mode even lies above the negative chemical potential, which
could be identified as the upper limit of the Bogoliubov spectrum.
Obviously, the variational ansatz with coupled Gaussians and spherical
harmonics is not as appropriate for the self-trapped monopolar
condensate as for the condensate without long-range interaction.  To
obtain still better results in the variational ansatz, it would be
necessary to use more than $N=6$ coupled Gaussians. This, however,
leads to numerical difficulties, since the matrix $\mathbf{K}$ in
the equations of motion (\ref{eq:tdvpeom}) becomes more and more
ill-conditioned when the number of Gaussians is increased, which leads
to an inaccurate solution of the linear system of equations.

\section{Conclusion and outlook}
\label{sec:conclusion-outlook}

We investigated the Bogoliubov spectrum of the self-trapped monopolar
condensate full-numerically with the finite-difference method. With
this method, we were able to calculate many modes for angular momenta
from $l=0$ to $l=6$. We found a similar structure as in the spectra
of alkali atoms. The behavior could be explained by
quantum defect theory, and it was found that practically the entire 
spectrum can be described by a simple Rydberg formula with quantum defects.

As an alternative to full-numerical calculations of condensate
excitations a new variational ansatz was presented which combines
coupled Gaussians with spherical harmonics.  Using the time-dependent
variational principle we could derive the equations of motion for the
variational parameters. We applied the variational ansatz to two
different systems.  For condensates with 
an attractive short-range interaction we found that 
there is a good agreement between the quantum mechanical eigenfrequencies
of the lowest Bogoliubov excitations and the eigenvalues of the Jacobian 
stability matrix. In this way we have been able to link 
the concepts of stability in quantum mechanics  and in classical dynamical 
systems with each other.

For self-trapped condensates
with additional $1/r$ interaction we also found a good agreement for
the very lowest modes, but the variational ansatz works less well
for higher modes.  What is the reason for this?  For
the condensate without long-range interaction in an external trap, the
confining radially symmetric harmonic potential 
dominates the properties of the system in a wide range of the
scattering length. The contact
interaction quasi acts as a perturbation.  Therefore, a
variational ansatz in which the radial part is determined by
Gaussians is very well adapted to describe the stationary solutions and
their excitations.

For the self-trapped monopolar condensate, on the other hand, an
external trap is missing and the interactions alone determine the
properties of the system. As pointed out in
Sec.~\ref{sec:quant-defect-analys}, the asymptotic form for $r \to
\infty$ of the BDG equations is equivalent to the Schr\"odinger
equation of the hydrogen atom. Therefore in that range the solutions
$u$ and $v$ could be approximated by Laguerre polynomials and the
exponential function $\exp(-\alpha r)$ with some $\alpha > 0$.  A
variational ansatz with coupled Gaussians and spherical harmonics is
not well suited to reproduce this asymptotic behavior. However, as
soon as a radially symmetric trap is switched on, the agreement
between the quantum mechanical and the nonlinear dynamics excitations
is present again also for the higher modes.

All together it was shown that especially in the case without
long-range interactions the extended variational ansatz works well and
can reproduce the lowest modes for arbitrary angular momenta, which is
significant progress compared to the ansatz with coupled Gaussians
only.  If one is interested only in the lowest modes, the ansatz is a
valid alternative to the full-numerical calculations.

So far, we only calculated the linear dynamics in the vicinity of a
stationary solution. It remains to be shown whether or not the ansatz
is capable of describing also the full nonlinear dynamics of a
BEC\@. Furthermore, the present ansatz is restricted to radially
symmetric systems. To calculate excitations of cylindrically symmetric
systems with arbitrary angular momenta, which would be necessary,
e.g., for condensates with dipole-dipole long-range interactions, an
extension of the ansatz is required. For dipolar condensates such an
ansatz would be of interest, since the dipolar interaction offers the
new possibility for a condensate to collapse with $m=2, 3, \dots$
symmetry, the so-called angular collapse \cite{Wilson09a}.

\begin{acknowledgments}
  This work was supported by Deutsche Forschungsgemeinschaft.
\end{acknowledgments}

\appendix*

\section{Integrals for the variational ansatz with coupled Gaussians
  and spherical harmonics}
\label{sec:integr-vari-ansatz}

We give the integrals necessary for setting up the equations of motion
resulting from the time-dependent variational principle for the new
variational ansatz Eq.~(\ref{eq:varansatzsph}). We need the matrix and
vector
\begin{subequations}
  \begin{align}
    K_{ij} &=
    \qmprod{\firstpderiv{\psi}{z_i}}{\firstpderiv{\psi}{z_j}}, \\
    h_i &= \qmprod{\firstpderiv{\psi}{z_i}}{\hat{H} \psi},
  \end{align}
\end{subequations}
where the mean-field Hamiltonian $\hat{H}$ consists of four parts
\begin{align}
  \hat{H} &= \hat{T} + V_\text{ext} + V_\text{s} + V_\text{m} \nonumber \\
  &= -\Delta + \gamma_r^2 r^2 + 8 \pi a \abs{\psi(\vec{r})}^2
  - 2 \int \dd^3 r^\prime
  \frac{\abs{\psi(\vec{r}^\prime)}^2}{\abs{\vec{r} - \vec{r}^\prime}}.
\end{align}

To calculate the integrals, we write the
ansatz~(\ref{eq:varansatzsph}) in a slightly different form
\begin{align}
  \psi = \sum\limits_{k=1}^N
  \sum\limits_{l,m}
  d_{lm}^k Y_{lm} (\theta,\phi) r^l
  \eto{-A_r^k r^2 - \gamma^k},
\end{align}
where all $d_{00}^k \equiv 1$ have to be treated as constants, and not
as variational parameters.

\subsection*{Integrals of the K matrix}

For the elements of the $\mathbf{K}$ matrix, one needs the integrals
over two spherical harmonics, which because of their orthogonality are
given by Kronecker deltas, and the integrals over the radial
coordinate, which are all of the form
\begin{align}
  I_r = \int\limits_0^\infty \dd r \,
  r^l \exp
  \left(
    -A r^2
  \right).
\end{align}
With the substitution $r \to t = A r^2$, one can use the Gamma
function \cite{Arfken05a} to write
\begin{align}
  \label{eq:radint}
  I_r = \frac{1}{2} A^{-(l+1)/2}
  \Gamma [(l+1)/2].
\end{align}
For the elements of the $\mathbf{K}$ matrix we then obtain, with the
definitions $A_r^{kl} \equiv A_r^k + (A_r^l)^*$ and $\gamma^{kl}
\equiv \gamma^k + (\gamma^l)^*$
\begin{align}
  \qmprod{\firstpderiv{\psi}{d_{l_2m_2}^l}}{\firstpderiv{\psi}{d_{l_1m_1}^k}}
  &=\frac{1}{2} \delta_{l_1l_2} \delta_{m_1m_2}
  \frac{\Gamma(l_1+3/2)}{(A_r^{kl})^{l_1+3/2}} \eto{-\gamma^{kl}}, \displaybreak[0]\\
  \qmprod{\firstpderiv{\psi}{d_{l_2m_2}^l}}{\firstpderiv{\psi}{A_r^k}}
  &=-\frac{1}{2} d_{l_2m_2}^k
  \frac{\Gamma(l_2+5/2)}{(A_r^{kl})^{l_2+5/2}} \eto{-\gamma^{kl}}, \displaybreak[0]\\
  \qmprod{\firstpderiv{\psi}{d_{l_2m_2}^l}}{\firstpderiv{\psi}{\gamma^k}}
  &=-\frac{1}{2} d_{l_2m_2}^k
  \frac{\Gamma(l_2+3/2)}{(A_r^{kl})^{l_2+3/2}} \eto{-\gamma^{kl}},
\end{align}
\begin{align}
  \qmprod{\firstpderiv{\psi}{A_r^l}}{\firstpderiv{\psi}{A_r^k}}
  &=\frac{1}{2} \sum\limits_{l_1,m_1}
  (d_{l_1m_1}^l)^* d_{l_1m_1}^k
  \frac{\Gamma(l_1+7/2)}{(A_r^{kl})^{l_1+7/2}} \eto{-\gamma^{kl}}, \displaybreak[0]\\
  \qmprod{\firstpderiv{\psi}{A_r^l}}{\firstpderiv{\psi}{\gamma^k}}
  &=\frac{1}{2} \sum\limits_{l_1,m_1}
  (d_{l_1m_1}^l)^* d_{l_1m_1}^k
  \frac{\Gamma(l_1+5/2)}{(A_r^{kl})^{l_1+5/2}} \eto{-\gamma^{kl}}, \displaybreak[0]\\
  \qmprod{\firstpderiv{\psi}{\gamma^l}}{\firstpderiv{\psi}{\gamma^k}}
  &=\frac{1}{2} \sum\limits_{l_1,m_1}
  (d_{l_1m_1}^l)^* d_{l_1m_1}^k
  \frac{\Gamma(l_1+3/2)}{(A_r^{kl})^{l_1+3/2}} \eto{-\gamma^{kl}}.
\end{align}

\subsection*{Integrals of the kinetic term}

For the calculation of the integrals of the kinetic term, one lets the
Laplacian act on the variational ansatz. The integrals of the
resulting terms can then be evaluated using Eq.~(\ref{eq:radint}),
which leads to
\begin{widetext}
  \begin{align}
    \qmprod{\firstpderiv{\psi}{d_{l_2m_2}^l}}{\hat{T} \psi}
    &= \frac{1}{2} \sum\limits_{k=1}^N d_{l_2m_2}^k
    \left[
      (4l_2+6) A_r^k \frac{\Gamma(l_2+3/2)}{(A_r^{kl})^{l_2+3/2}}
      - 4 (A_r^k)^2 \frac{\Gamma(l_2+5/2)}{(A_r^{kl})^{l_2+5/2}}
    \right] \eto{-\gamma^{kl}}, \\
    \qmprod{\firstpderiv{\psi}{A_r^l}}{\hat{T} \psi}
    &= -\frac{1}{2} \sum\limits_{k=1}^N \sum\limits_{l_1,m_1}
    (d_{l_1m_1}^l)^* d_{l_1m_1}^k
    \left[
      (4l_1+6) A_r^k \frac{\Gamma(l_1+5/2)}{(A_r^{kl})^{l_1+5/2}}
      - 4 (A_r^k)^2 \frac{\Gamma(l_1+7/2)}{(A_r^{kl})^{l_1+7/2}}
    \right] \eto{-\gamma^{kl}}, \\
    \qmprod{\firstpderiv{\psi}{\gamma^l}}{\hat{T} \psi}
    &= -\frac{1}{2} \sum\limits_{k=1}^N \sum\limits_{l_1,m_1}
    (d_{l_1m_1}^l)^* d_{l_1m_1}^k
    \left[
      (4l_1+6) A_r^k \frac{\Gamma(l_1+3/2)}{(A_r^{kl})^{l_1+3/2}}
      - 4 (A_r^k)^2 \frac{\Gamma(l_1+5/2)}{(A_r^{kl})^{l_1+5/2}}
    \right] \eto{-\gamma^{kl}}.
  \end{align}
\end{widetext}

\subsection*{Integrals of the trapping potential}

The integrals for the trapping potential are straightforward:
\begin{align}
  \qmprod{\firstpderiv{\psi}{d_{l_2m_2}^l}}{V_\text{ext} \psi}
  &= \frac{1}{2} \gamma_r^2
  \sum\limits_{k=1}^N d_{l_2m_2}^k
  \frac{\Gamma(l_2+5/2)}{(A_r^{kl})^{l_2+5/2}}
  \eto{-\gamma^{kl}}, \displaybreak[0] \\
  \qmprod{\firstpderiv{\psi}{A_r^l}}{V_\text{ext} \psi}
  &= -\frac{1}{2} \gamma_r^2
  \sum\limits_{k=1}^N \sum\limits_{l_1,m_1}
  (d_{l_1m_1}^l)^* d_{l_1m_1}^k \nonumber \\
  &\times \frac{\Gamma(l_1+7/2)}{(A_r^{kl})^{l_1+7/2}}
  \eto{-\gamma^{kl}}, \displaybreak[0] \\
\qmprod{\firstpderiv{\psi}{\gamma^l}}{V_\text{ext} \psi}
  &= -\frac{1}{2} \gamma_r^2
  \sum\limits_{k=1}^N \sum\limits_{l_1,m_1}
  (d_{l_1m_1}^l)^* d_{l_1m_1}^k \nonumber \\
  &\times \frac{\Gamma(l_1+5/2)}{(A_r^{kl})^{l_1+5/2}}
  \eto{-\gamma^{kl}}.
\end{align}

\subsection*{Integrals of the scattering term}

To write down the integrals of the scattering term, we introduce the
new abbreviations $A_r^{ijkl} = A_r^{ij} + A_r^{kl}$, $\gamma^{ijkl} =
\gamma^{ij} + \gamma^{kl}$, and for the integral over four spherical
harmonics the notation
\begin{align}
  \label{eq:intfoursph}
  &I_\Omega^ {(4)} (l_1,m_1;l_2,m_2;l_3,m_3;l_4,m_4) \nonumber \\
  &= \int \dd \Omega \,
  Y_{l_1m_1}(\theta,\phi) Y_{l_2m_3}(\theta,\phi)
  Y_{l_3m_3}(\theta,\phi) Y_{l_4m_4}(\theta,\phi),
\end{align}
where $\dd \Omega = \dd \phi \, \dd \theta \, \sin\theta$ is the 
differential solid
angle element of the angular coordinates. Using again
Eq.~(\ref{eq:radint}), we obtain for the integrals
\begin{widetext}
  \begin{align}
    \qmprod{\firstpderiv{\psi}{d_{l_2m_2}^l}}{V_\text{s} \psi}
    &= 4 \pi a \sum_{i,j,k=1}^N \sum_{l_1,m_1} \sum_{l_3,m_3} \sum_{l_4,m_4}
    (-1)^{m_2+m_4} (d_{l_4m_4}^j)^* d_{l_3m_3}^i d_{l_1m_1}^k
    \frac{\Gamma[(l_1+l_2+l_3+l_4+3)/2]}{(A_r^{ijkl})^{-(l_1+l_2+l_3+l_4+3)/2}}
    \nonumber \\
    &\times I_\Omega^ {(4)} (l_2,-m_2;l_4,-m_4;l_3,m_3;l_1,m_1)
    \eto{-\gamma^{ijkl}}, \\
    \qmprod{\firstpderiv{\psi}{A_r^l}}{V_\text{s} \psi}
    &= -4 \pi a \sum_{i,j,k=1}^N \sum_{l_1,m_1} \sum_{l_2,m_2}
    \sum_{l_3,m_3} \sum_{l_4,m_4}
    (-1)^{m_2+m_4} (d_{l_2m_2}^l)^* (d_{l_4m_4}^j)^* d_{l_3m_3}^i d_{l_1m_1}^k
    \frac{\Gamma[(l_1+l_2+l_3+l_4+5)/2]}{(A_r^{ijkl})^{-(l_1+l_2+l_3+l_4+5)/2}}
    \nonumber \\
    &\times I_\Omega^ {(4)} (l_2,-m_2;l_4,-m_4;l_3,m_3;l_1,m_1)
    \eto{-\gamma^{ijkl}}, \\
    \qmprod{\firstpderiv{\psi}{\gamma^l}}{V_\text{s} \psi}
    &= -4 \pi a \sum_{i,j,k=1}^N \sum_{l_1,m_1} \sum_{l_2,m_2}
    \sum_{l_3,m_3} \sum_{l_4,m_4}
    (-1)^{m_2+m_4} (d_{l_2m_2}^l)^* (d_{l_4m_4}^j)^* d_{l_3m_3}^i d_{l_1m_1}^k
    \frac{\Gamma[(l_1+l_2+l_3+l_4+3)/2]}{(A_r^{ijkl})^{-(l_1+l_2+l_3+l_4+3)/2}}
    \nonumber \\
    &\times I_\Omega^ {(4)} (l_2,-m_2;l_4,-m_4;l_3,m_3;l_1,m_1)
    \eto{-\gamma^{ijkl}}.
  \end{align}
\end{widetext}
An analytical expression for $I_\Omega^{(4)}$ is found by noting that
the product of two spherical harmonics can be expressed by a series of
spherical harmonics
\begin{align}
  Y_{l_1m_1}(\theta,\phi) Y_{l_2m_2}(\theta,\phi) =
  \sum\limits_{l=0}^\infty \sum\limits_{m=-l}^l
  C_l^m{}_{l_1}^{m_1}{}_{l_2}^{m_2} Y_{lm}(\theta,\phi),
\end{align}
where the coefficients $C_l^m{}_{l_1}^{m_1}{}_{l_2}^{m_2}$ can be
written in terms of Wigner 3j symbols \cite{Thompson94a}
\begin{align}
  \label{eq:intsphcoef}
  C_l^m{}_{l_1}^{m_1}{}_{l_2}^{m_2} &= (-1)^m
  \sqrt{\frac{(2l_1+1)(2l_2+1)(2l+1)}{4\pi}} \nonumber \\
  &\times
  \begin{pmatrix}
    l_1 & l_2 & l \\
    0 & 0 & 0
  \end{pmatrix}
  \begin{pmatrix}
    l_1 & l_2 & l \\
    m_1 & m_2 & -m
  \end{pmatrix}.
\end{align}
Applying this expansion twice in the integral
Eq.~(\ref{eq:intfoursph}), we obtain
\begin{align}
  &I_\Omega^ {(4)} (l_1,m_1;l_2,m_2;l_3,m_3;l_4,m_4) \nonumber \\
  &= \sum\limits_{l=0}^\infty \sum\limits_{m=-l}^l
  (-1)^m C_l^m{}_{l_1}^{m_1}{}_{l_2}^{m_2}
  C_{\hphantom{-}l}^{-m}{}_{l_3}^{m_3}{}_{l_4}^{m_4}.
\end{align}
The infinite sum can be cut off, since a Wigner 3j symbol is zero, if
the triangle inequalities $\abs{l_1-l_2} \leq l \leq l_1+l_2$ or
$\abs{l_3-l_4} \leq l \leq l_3+l_4$ are not fulfilled, and
$l_1,\dots,l_4$ cannot be greater than the largest angular momentum
included in the variational ansatz.

\subsection*{Integrals of the monopolar term}

The integrals for the monopolar term read
\begin{widetext}
  \begin{align}
    \qmprod{\firstpderiv{\psi}{d_{l_2m_2}^l}}{V_\text{m} \psi}
    &= -2 \sum_{i,j,k=1}^N \sum_{l_1,m_1} \sum_{l_3,m_3} \sum_{l_4,m_4}
    (d_{l_4m_4}^j)^* d_{l_3m_3}^i d_{l_1m_1}^k I_{\text{m},0}, \\
    \qmprod{\firstpderiv{\psi}{A_r^l}}{V_\text{m} \psi}
    &= 2 \sum_{i,j,k=1}^N \sum_{l_1,m_1} \sum_{l_2,m_2} \sum_{l_3,m_3} \sum_{l_4,m_4}
    (d_{l_2m_2}^l)^* (d_{l_4m_4}^j)^* d_{l_3m_3}^i d_{l_1m_1}^k I_{\text{m},2}, \\
    \qmprod{\firstpderiv{\psi}{\gamma^l}}{V_\text{m} \psi}
    &= 2 \sum_{i,j,k=1}^N \sum_{l_1,m_1} \sum_{l_2,m_2} \sum_{l_3,m_3} \sum_{l_4,m_4}
    (d_{l_2m_2}^l)^* (d_{l_4m_4}^j)^* d_{l_3m_3}^i d_{l_1m_1}^k I_{\text{m},0},
  \end{align}
  with the definition
  \begin{align}
    I_{\text{m},p} =
    \int \dd \Omega \int\limits_0^\infty \dd r
    \int \dd \Omega^\prime \int\limits_0^\infty \dd r^\prime \,
    &\frac{1}{\abs{\vec{r}-\vec{r}^\prime}} 
    Y_{l_2m_2}^*(\theta,\phi) Y_{l_1m_1}(\theta,\phi)
    Y_{l_4m_4}^*(\theta^\prime,\phi^\prime)
    Y_{l_3m_3}(\theta^\prime,\phi^\prime) \nonumber \\
    &\times r^{l_1+l_2+p+2} (r^\prime)^{l_3+l_4+2}
    \eto{-A_r^{kl} r^2} \eto{-A_r^{ij} (r^\prime)^2}.
  \end{align}
\end{widetext}
To calculate this integral, the monopolar interaction potential
$1/\abs{\vec{r}-\vec{r}^\prime}$ is expanded in terms of multipoles
\cite{Arfken05a}
\begin{align}
  \label{eq:multipole}
  \frac{1}{\abs{\vec{r}-\vec{r}^\prime}} =
  \sum\limits_{l=0}^\infty \sum\limits_{m=-l}^l
  \frac{4\pi}{2l+1} \frac{r_<^l}{r_>^{l+1}}
  Y_{lm}^*(\theta,\phi) Y_{lm}(\theta^\prime,\phi^\prime).
\end{align}
The integral $I_{\text{m},p}$ then separates into two integrals over
the angular coordinates $\Omega,\Omega^\prime$, which can be expressed
with the coefficients $C_l^m{}_{l_1}^{m_1}{}_{l_2}^{m_2}$ from
Eq.~(\ref{eq:intsphcoef}), and one integral over the radial
coordinates $r,r^\prime$, which is of Gaussian type and can be solved
analytically. For $I_{\text{m},p}$ we obtain
\begin{align}
  \label{eq:intmonores}
  I_{\text{m},p} =
  \sum\limits_{l=0}^\infty \sum\limits_{m=-l}^l
  \frac{4\pi}{2l+1} I_\text{m}^\Omega I_\text{m}^{\Omega^\prime}
  I_{\text{m},p}^{r,r^\prime},
\end{align}
with the individual integrals
\begin{align}
  I_\text{m}^\Omega &= (-1)^m
  C_{l_2}^{m_2}{}_{\hphantom{-}l}^{-m}{}_{l_1}^{m_1}, \\
  I_\text{m}^{\Omega^\prime} &=
  C_{l_4}^{m_4}{}_{l}^{m}{}_{l_3}^{m_3},
\end{align}
and
\begin{widetext}
  \begin{align}
    I_{\text{m},p}^{r,r^\prime} =
    &\frac{1}{4} \frac{[(l_3+l_4-l)/2]!}{
      (A_r^{ij})^{(l_3+l_4-l+2)/2} (A_r^{ijkl})^{(l_1+l_2+l+p+3)/2}}
    \sum\limits_{\alpha=0}^{\frac{l_3+l_4-l}{2}} \frac{1}{\alpha!}
    \left(
      \frac{A_r^{ij}}{A_r^{ijkl}}
    \right)^\alpha
    \Gamma[(l_1+l_2+l+p+2\alpha+3)/2] \nonumber \\
    + &\frac{1}{4} \frac{[(l_1+l_2-l+p)/2]!}{
      (A_r^{kl})^{(l_1+l_2-l+p+2)/2} (A_r^{ijkl})^{(l_3+l_4+l+3)/2}}
    \sum\limits_{\alpha=0}^{\frac{l_1+l_2-l+p}{2}} \frac{1}{\alpha!}
    \left(
      \frac{A_r^{kl}}{A_r^{ijkl}}
    \right)^\alpha
    \Gamma[(l_3+l_4+l+2\alpha+3)/2].
  \end{align}
\end{widetext}
The infinite sum in Eq.~(\ref{eq:intmonores}) can be cut off again due
to the properties of the Wigner 3j symbols. Thus
all integrals necessary for setting up the equations of motion for the
variational parameters for the ansatz with coupled Gaussians and spherical
harmonics have been calculated analytically.

\end{document}